\documentclass[twocolumn,preprint]{aastex62}
\usepackage[T1]{fontenc}
\usepackage[english]{babel}
\usepackage{natbib}
\usepackage{ulem,xcolor}
\usepackage{amsmath, graphicx}
\usepackage{soul}
\def\mw{{microwave}}

\def\gs{{gyrosynchrotron}}

\newcommand{\gf}{}

\begin{document}

\title{Spectral evolution of mildly relativistic electrons in solar flares}

\author[0000-0001-5557-2100]{Gregory D. Fleishman}
	\affil{Center For Solar-Terrestrial Research, New Jersey Institute of Technology, Newark, NJ 07102, USA}

\affil{Institut f\"ur Sonnenphysik (KIS), Georges-Köhler-Allee 401 A, D-79110 Freiburg, Germany}

    \author[0000-0003-0366-1851]{Tatyana Kaltman }
    \affil{Institut f\"ur Sonnenphysik (KIS), Georges-Köhler-Allee 401 A, D-79110 Freiburg, Germany}

    \author[0000-0003-2872-2614]{Sijie Yu} 
	\affil{Center For Solar-Terrestrial Research, New Jersey Institute of Technology, Newark, NJ 07102, USA}

    \author[0000-0003-2846-2453]{Gelu Nita}
    \affil{Center For Solar-Terrestrial Research, New Jersey Institute of Technology, Newark, NJ 07102, USA}


\begin{abstract}
Solar flares are a powerful engine capable of accelerating ambient plasma particles to high nonthermal energies. A key characteristic of these particles is their energy spectrum. Numerous studies of nonthermal hard X-ray emission established that a typical spectral evolution of the nonthermal electrons with energies of dozens keV follows a soft-hard-soft pattern in impulsive events and soft-hard-harder pattern in some long-duration events. Here we extend the study of the spectral evolution to the case of flare-accelerated mildly relativistic electrons primarily responsible for the flare microwave emission. We examine the spatially resolved \mw\ emission observed with the Expanded Owens Valley Solar Array in twelve solar flares and find that they also typically follow a soft-hard-soft spectral evolution. However, the revealed spectral evolution of the mildly relativistic electrons covers the spectral index range from 2-4 to 15 or more and, thus, is much more prominent than for the X-ray-producing electrons. The evolution of the spectral index is closely correlated with the emission flux and does not show any noticeable time delay. The spectral index displays a prominent correlation with the brightness temperature with a pattern similar to all events considered. We conclude that the revealed relationships and the evolutionary pattern is an inherent property of particle acceleration in the solar flares. 
\end{abstract}


\section{Introduction}
Solar flares are, perhaps, the most powerful factories in our solar system that are capable of accelerating large amounts of charged particles to high energies strongly exceeding the mean energies of the ambient plasma particles \citep[see, e.g., the review by][]{2017LRSP...14....2B}. These high energy particles manifest themselves in transient electromagnetic emission that gives a major contribution to the microwave, X-ray, and $\gamma$-ray spectra of solar flares. Understanding the flare acceleration process requires detailed analysis of those spectra as properties of the acceleration engine itself are encoded in the observational characteristics of the nonthermal emissions \citep{1979ApJ...227.1072B}. In particular, it has been recognized that evolution of the observed spectra imposes highly important constraints on the particle acceleration \citep{1983ApJ...267..433B,1991A&A...245..271A,1992ApJ...398..350H,1996ApJ...461..445M,1997JGR...10214631M,2004ApJ...610..550P,2006A&A...458..641G,Byk_Fl_2009,2016ApJ...830...28P,2017ApJ...835..124E}.

Observationally, spectral evolution of nonthermal emission in solar flares has been studied over several decades, primarily, in the hard X-ray (HXR) domain, where the photon spectra fall with energy. Often, such spectra have been studied using spectral model fitting that assumes a single or a broken power-law in the photon or electron energy domain \citep{2011SSRv..159..107H,2011SSRv..159..301K}. In the case of the broken power-laws, spectra with both break up and break down were reported \citep[see, e.g.,][and references therein]{1998A&A...334.1099T, 2011SSRv..159..225W}. \cite{1985SoPh..100..465D,1988SoPh..118...49D} reported a statistical study of spectral properties of solar flares recorded by Solar Maximum Mission within the context of flare categorization onto impulsive and gradual/microwave-rich following the classification proposed by \citet{1983ASSL..102..307T,1983SoPh...86...91T}. According to \citet{1988ApJ...324.1118K}, a vast majority ($\sim81\%$) of solar flares are impulsive, $\sim15\%$ are accompanied by a \mw\ burst with gradual rise and fall, $\sim3\%$ are gradual, while remaining are mainly thermal.
\cite{1985SoPh..100..465D,1988SoPh..118...49D} found that impulsive flares typically demonstrate a soft-hard-soft (SHS) spectral evolution during the rise-peak-decay X-ray flux evolution, which was attributed to the acceleration process itself. In contrast, gradual flares often show a soft-hard-harder (SHH) pattern with overall harder (flatter) HXR spectra compared with the impulsive ones \citep[see also][]{2008ApJ...683.1180G}. This behavior was interpreted as a result of nonthermal electron trapping in large coronal loops or a second-stage acceleration \citep[e.g.,][]{1979ApJ...227.1072B}. 

\citet{2004A&A...426.1093G} extended the finding of \citet{1985SoPh..100..465D} about the SHS behavior of the impulsive flares to flares with multiple temporal peaks. They demonstrated that the SHS pattern relates not only to the flare as a whole, but also to each individual temporal peak; thus, strengthening the case that the SHS spectral evolution is an inherent property of the flare acceleration process. Numerous studies confirmed these findings down to very short, even sub-second, emission episodes \citep[see, e.g.,][]{2018ApJ...867...84G,2024A&A...684A.215C}. 

Joint analysis of X-ray and \mw\ observations has a potential of much better constraining the nonthermal particle populations than X-ray data alone \citep[see, e.g.,][]{2000ApJ...545.1116S,2011SSRv..159..225W, 2021ApJ...915L...1D,2025AN....34640134V}; however, it revealed a lot of controversy. \citet{2000ApJ...545.1116S} analyzed 27 flares (23 impulsive and four non-impulsive/gradual) and found that the energy spectral indices of nonthermal electrons, $\delta_X$ and $\delta_r$, defined from the X-ray  and \mw\ spectral fits, respectively, do not agree with each other. In the case of gradual flares, the difference reported is about $\delta_X-\delta_r\approx2$. And even for the impulsive flare, when the X-ray and \mw\ light curves are very similar, \citet{2000ApJ...545.1116S} reported the difference of $\delta_X-\delta_r\approx0.8$; thus, their reported $\delta_r$ were systematically lower than $\delta_X$, i.e., the \mw-inferred electron spectra systematically flatter/harder than the X-ray-inferred ones \citep[see also][]{2025AN....34640134V}. 
It was recognized that the revealed difference, at least in part, can be due to different locations of the X-ray and \mw\ sources: the HXR emission is mainly produced by precipitating electrons in dense chromospheric footpoints, while the \mw\ emission---primarily from coronal sources. 

\citet{2010ApJ...714.1108K} removed this ambiguity by studying a limb flare, 2007-Dec-31, whose footpoints were occulted; thus, the observed HXR emission came from a coronal thin-target source spatially coincident with the \mw\ source. Yet, they reported a \mw-inferred spectral index to be systematically smaller than the HXR-inferred one, $\delta_X-\delta_r\approx0.3$, while closely correlated in time at the course of the bursts. It is important to realize that this and many other reports first (1) computed the slope $\alpha$ of the \mw\ spectrum in the optically thin domain and then (2) employed an overly simplistic Dulk-Marsh approximation to convert it to $\delta_r$, which is inaccurate for quantitative comparisons. \citet{Fl_Kuzn_2010} employed newly developed fast gyrosynchrotron codes, which are efficient and precise, to  demonstrate that the \gs\ spectrum computed using the HXR-inferred source parameters perfectly matches the observed \mw\ spectrum in the 2007-Dec-31 flare. This eliminated the inconsistency reported by \citet{2010ApJ...714.1108K} and emphasized the importance of using accurate treatment of the \mw\ emission in quantitative studies. 


Although some inconsistencies between the HXR- and \mw-inferred spectral diagnostics can be ascribed to the overly simplistic treatment discussed above, the joint X-ray and \mw\ analysis does reveal real differences in the spectral properties of the nonthermal electrons. {\gf This, at least to a certain extent, can be due to different electron energies sampled by HXR and \mw\ emissions \citep{1998A&A...334.1099T} or different spatial populations of the accelerated electrons}.
Indeed, even impulsive flares that show a classical SHS spectral behavior in the X-ray domain, often show a SHH pattern in the \mw\ domain \citep{2002ESASP.506..339M,2007PASJ...59..373N,2007ApJ...671L.197N,2009SoPh..257..335N}. This phenomenon was broadly interpreted as an effect of nonthermal electron trapping in coronal flaring loops \citep{1994R&QE...37..557M,1998ARA&A..36..131B,2000ApJ...531.1109L,2002ApJ...580L.185M,2023ApJ...953..174F}.

However, \mw\ bursts offer a noticeably richer spectral evolution patterns than the X-ray emission. For example, \citet{2023ApJ...954..122L} found that a majority of `cold' flares they studied display a hard-soft-hard (HSH) spectral evolution, see also \citet{2000ApJ...543..457L},---just the opposite to the X-ray bursts. \citet{2023ApJ...954..122L} noted that the spectral evolution of the \mw\ emission does not necessarily reproduce the spectral evolution of the emitting nonthermal electrons for several reasons; for example, thermal free-free emission from ambient coronal plasma or evolving pitch-angle anisotropy may affect the \mw\ spectral slope. Spatially resolved observations sometimes reveal a distinct thermal source at the burst decay phase \citep{2024SoPh..299...80S}, which can cause hardening of the total power \mw\ spectrum. 

An important distinction between the HXR and \mw\ emission is that the HXR emission is optically thin, while the \mw\ spectra typically include both optically thin and thick regions. Moreover, while the HXR spectral shape is primarily determined by the energy shape (slope) of the nonthermal electron spectrum \citep[although several other physical processes affect the HXR properties as well; see, e.g.,][]{2011SSRv..159..107H,2011SSRv..159..301K},  the \mw\ spectrum (including the optically thin slope) is defined not only by the slope and extent of the electron distribution over energy. It also noticeably depends on the angular distribution of the nonthermal electrons, the magnetic field strength, the viewing angle, and the ambient thermal number density \citep[see, e.g.,][]{Fl_Meln_2003b}. It is further important, but was typically ignored, that the energy spectral index $\delta$ of the nonthermal electron distribution affects not only the optically thin, but also the optically thick part of the \mw\ spectrum \citep[see, e.g., supplemental videos in][]{2020Sci...367..278F,2022Natur.606..674F}. 

This implies that a meaningful interpretation of the \mw\ data, including comparisons with HXR, requires a more rigorous analysis than has often been performed using a simplified treatment of the optically thin portion of the \mw\ spectrum. Such analysis has been performed for several case studies, which overall suggest that the same population of the nonthermal electrons is often responsible for both \mw\ and HXR emissions, like in the already mentioned occulted 2007-Dec-31 flare. In some cases this distribution follows a single power-law in the entire accessible energy range of the nonthermal electrons; in others it can deviate from such power-laws and show break ups or downs. 

\citet{Fl_etal_2011} found that, in a 'cold' 2002-Jul-30 solar flare, a single power-law spectrum with $\delta\approx3.5$ inferred from HXR spectral fitting above 6\,keV is consistent with the \mw\ data over the entire flare duration. In this event, the \mw\ emission was dominated by a contribution from the very acceleration region with no signature of electron trapping; no significant spectral evolution has been detected in this event. In contrast, a 2002-Apr-11 flare displayed contributions from both the acceleration region during the impulsive phase and a trapped component during the decay phase \citep{Fl_etal_2013}. Again, a single power-law distribution with the same spectral index was consistent with both HXR and \mw\ data during the impulsive phase, while in the decay phase a  component trapped in a looptop of a flaring loop displayed hardening similar to that reported in other studies \citep[e.g.,][and many others]{1994R&QE...37..557M,1998ARA&A..36..131B,2000ApJ...531.1109L,2002ApJ...580L.185M,2007PASJ...59..373N,2007ApJ...671L.197N,2009SoPh..257..335N}.

To quantitatively model emissions from such flaring loops, \citet{Nita_etal_2015} developed a powerful modeling tool, called GX Simulator, that permitted developing 3D data-constrained models, whose synthesized emissions are to be directly compared with data for model fine-tuning and validation \citep[see also][]{2023ApJS..267....6N}. Several studies employed this tool to create 3D models that reveal relationships of the nonthermal electron populations with HXR and \mw\ emissions they generate. For example, \citet{Fl_Xu_etal_2016} found that a single broken power-law electron distribution with a break up around 40\,keV located in a single dense coronal loop is fully consistent with both imaging and spectral HXR and \mw\ observations of a 'coronal-thick-target' 2002-Apr-12 flare. {\gf Similar break-up spectra were inferred from multiwavelength radio/mm/HXR observations of several flares, where the radio- and HXR-derived spectral indices converged only above a break energy of several hundred keV
\citep{1998A&A...334.1099T,2021ApJ...915L...1D}.}

In contrast, a cold 2002-Mar-10 flare, whose HXR spectrum measured in the 20--1000\,keV range implied an electron distribution that softened with energy (a broken power-law with a break down at about 150\,keV), included two flux tubes with highly contrasting sizes and ambient densities. 
\citet{Fl_etal_2016} found that this HXR-derived electron population being appropriately partitioned between these two flaring loops is fully consistent with the available \mw\ constraints. 
\citet{2018ApJ...852...32K} considered a more complicated case of a 2015-Jun-22 M-class flare. They developed a sophisticated 3D model and performed joint analysis in the X-ray and \mw\ domains to conclude that at least four distinct coronal loops were involved. Those loops hosted different portions of the nonthermal flare electrons with different spectral properties, demonstrating a powerful complementarity of the X-ray and \mw\ diagnostics and highlighting the importance of spatially resolved observations, 
since the spectral properties--including the spectral evolution--can differ substantially across flare locations and among individual flaring loops. In the \mw\ domain, the capacity of measuring spatially resolved flare parameters comes with the use of imaging spectroscopy data \citep[employed, e.g., in][]{2020Sci...367..278F,2020NatAs...4.1140C,2022Natur.606..674F,2026NatAs..10..363F} available from the Expanded Owens Valley Solar Array \citep[EOVSA,][]{Gary_etal_2018}. 

\citet{2021ApJ...908L..55C} performed a joint spectral fitting of HXR and \mw\ data from a spatially resolved coronal source in the rise phase of the 2017-Sep-10 eruptive flare \citep{Gary_etal_2018} and concluded that the data cannot be simultaneously fit using a single power-law spectrum. Instead, the data suggest a break down spectrum with the break energy above $\sim160$\,keV similar to that reported by \citet{Fl_etal_2016}. However, the most intriguing property of this joint fit is a contrasting evolution of the electron energy spectrum below and above the spectral break. While the low-energy spectral index varied around $\sim3.6$, the high-energy one displayed a remarkable hardening characterized by a spectral index $\delta$ decreasing from $>20$ to about 6 during 20\,s of the flare evolution. This implies that the \mw-producing mildly relativistic electrons can experience a much more prominent spectral evolution compared with the HXR-producing deka-keV ones. 

In this paper we investigate spectral evolution of mildly relativistic electrons derived from sequential pixel-by-pixel spectral model fitting of the \mw\ imaging spectroscopy data for twelve solar flares observed with EOVSA. We found that (1) the electron spectral index $\delta$ can vary in a given flare within very broad ranges, from $\sim2-3$ to $\gtrsim15$ during the course of the flare; thus, confirming \citet{2021ApJ...908L..55C} findings; (2) typically, the spectral evolution reveals SHS patterns; (3) this SHS evolution pertains to each (or most of the) temporal peaks of the flares having more than one individual peak; (4) there are rare cases that do not show substantial softening during the decay phase; and (5) the spectral index demonstrates a significant correlation with the brightness temperature reproducible across all the events considered.

\begin{table*}[ht]
\caption{Summary of EOVSA observations and analyzed solar flares.\tablenotemark{*} 
}
\begin{tabular}{l l l l | l l l l l l | l l l | l}
\hline\hline
  & \multicolumn{3}{c|}{General} 
 & \multicolumn{6}{c|}{EOVSA}
 & \multicolumn{3}{c|}{Morphology}
  & \multicolumn{1}{c}{Ref} \\
 \hline
\# & Date & GOES & Dur 
& $N_{\mathrm{ant}}$ & $N_{\mathrm{spw}}$ 
& $t_{\mathrm{peak}}$ 
& $S_{\mathrm{peak}}$ & $f_{\mathrm{peak}}$ & $t_{\mathrm{int}}$
& LCT & SFM & SpP
 \\
 &  &  & (min)
 &  &  
 & (UT)
 & (sfu) & (GHz) & (s) & 
 &  &  
 & \\
\hline
1  & 2017-Sep-07 & C4.5 & 0.3 & 13 & 30 & 18:41:39 & 153.4  & 9.48  & 1 & SP  & S   & SHS & [1] \\
2  & 2021-May-07 & M3.9 & 15  & 12 & 48 & 19:02:50 & 98.6   & 3.68  & 4 & MP  & S/D & SHS & [2] \\
3  & 2022-Mar-30 & X1.4 & 2   & 12 & 50 & 17:31:09 & 454.4  & 6.97  & 1 & QPP & D   & SHS &     \\
4  & 2022-Sep-16 & M6.7 & 2   & 11 & 50 & 15:51:05 & 215.6  & 6.24  & 4 & SP  & S   & SHS &    \\
5  & 2022-Oct-02 & X1.1 & 12  & 12 & 46 & 20:23:07 & 3831.7 & 17.29 & 8 & SP  & S/D & SHS & [3] \\
6  & 2024-Mar-08 & M1.3 & 2   & 10 & 40 & 21:24:43 & 172.3  & 5.75  & 4 & SP  & S   & SHS &     \\
7  & 2024-May-09 & X1.1 & 10  & 11 & 31 & 17:32:40 & 792.9  & 9.49  & 4 & MP  & S   & SHS &     \\
8  & 2024-May-11 & X5.8 & 14  & 9  & 36 & 01:17:31 & 2552.6 & 14.77 & 4 & MP  & S   & SHS &     \\
9  & 2024-May-12 & X1.0 & 4   & 12 & 37 & 16:19:29 & 1116.8 & 11.68 & 4 & SP  & S   & SHS &     \\
10 & 2024-May-14 & X8.8 & 4.5 & 11 & 37 & 16:47:18 & 7278.0 & 16.48 & 2 & DP  & D   & SH  &     \\
11 & 2024-Oct-31 & X2.0 & 5   & 11 & 50 & 21:17:39 & 1911.9 & 5.67  & 2 & MP  & S   & SH  &     \\
12 & 2024-Dec-08 & M1.5 & 3   & 12 & 30 & 16:08:27 & 276.9  & 6.16  & 2  & SP & S   & SHS &     \\
\hline
\end{tabular}
\tablenotetext{*}{
Dur = duration of the analyzed burst; 
$N_{\mathrm{ant}}$ = number of antennas; 
$N_{\mathrm{spw}}$ = number of spectral windows; 
$t_{\mathrm{peak}}$ = time of peak flux; 
$S_{\mathrm{peak}}$ = microwave peak flux density; 
$f_{\mathrm{peak}}$ = peak frequency; 
$t_{\mathrm{int}}$ = integration time. 
LCT = Light Curve Type (SP/DP/MP for single/double/multiple peaks, QPPs for quasi-periodic pulsations); 
SFM = Spectral Fitting Model (S/D for single or double source(s) on the LOS); 
SpP = Spectral Pattern (SH/SHS for soft-hard or soft-hard-soft evolution). 
References: [1] \citet{2025ApJ...988..260F}; [2] \citet{2026A&A...707A.158K}; [3] \citet{2026ApJ...999..179F}.
}
\label{table_combined_corr}
\end{table*}

\section{Observations and Data Set}
\subsection{Expanded Owens Valley Solar Array (EOVSA)}
\label{S_EOVSA}

The Expanded Owens Valley Solar Array (EOVSA; \citealt{2018ApJ...863...83G}) provides solar imaging spectroscopy at a cadence of 1\,s across 451 frequencies spanning 1–18 GHz. For all 12 flares analyzed in this study, raw visibility data were retrieved from the public EOVSA archive and calibrated using the standard processing pipeline, including delay, bandpass, and complex gain corrections.

Antenna gains were further refined through self-calibration \citep{1999Cornwell}, performed using the brightest compact emission near the burst peak for each event. The resulting calibration solutions were applied to the corresponding time interval analyzed for that flare.

Imaging was carried out using CLEAN algorithm with a circular restoring beam. For frequencies below 12\,GHz, the beam full width at half maximum (FWHM) was scaled as $60''\,f^{-1}$, where $f$ is the observing frequency in GHz. At higher frequencies, the FWHM was fixed at $5''$ to ensure stable sampling relative to the adopted image pixel size and to avoid artificial super-resolution effects.

Absolute flux calibration was performed by scaling the image-integrated flux densities to match independent single-dish total power measurements at each frequency band. For each reconstructed map, the root-mean-square (rms) brightness was computed and used as an estimate of the map uncertainty. In some cases, especially when spectra showed evidence of source complexity, we additionally searched for outliers in the spatially resolved spectra to estimate and account for the systematic errors. The \mw\ imaging spectroscopy data are prepared as four-dimensional (4D) data cubes that contain evolving (time dimension) multifrequency (frequency dimension) 2D images (two spatial dimensions).

These 4D data cubes can be viewed in various dimensions: the spatial domain, i.e., the images at different frequencies and times; the time domain, i.e., the light curves at different frequencies and locations; and the spectral domain, i.e., the spectra at different locations and times.
This study focuses on the spectral domain. Indeed, unlike the source properties in other mentioned domains, the shape of the \mw\ spectra is theoretically well understood, which permits, at least in principle, the derivation of physical parameters of the \mw\ source from analysis of such spectra \citep{Fl_etal_2009,Gary_etal_2013,2026ApJ...999..179F}.


Event-specific observational parameters, including the number of operational antennas $N_{\rm ant}$, the number of spectral windows $N_{\rm spw}$ used for imaging, peak flux density $S_{\rm peak}$, peak frequency $f_{\rm peak}$, and temporal integration time $t_{\rm int}$, are summarized in Table~\ref{table_combined_corr}.

\subsection{Overview of the Flare List}

\begin{figure*}\centering
\includegraphics[width=0.8\linewidth]{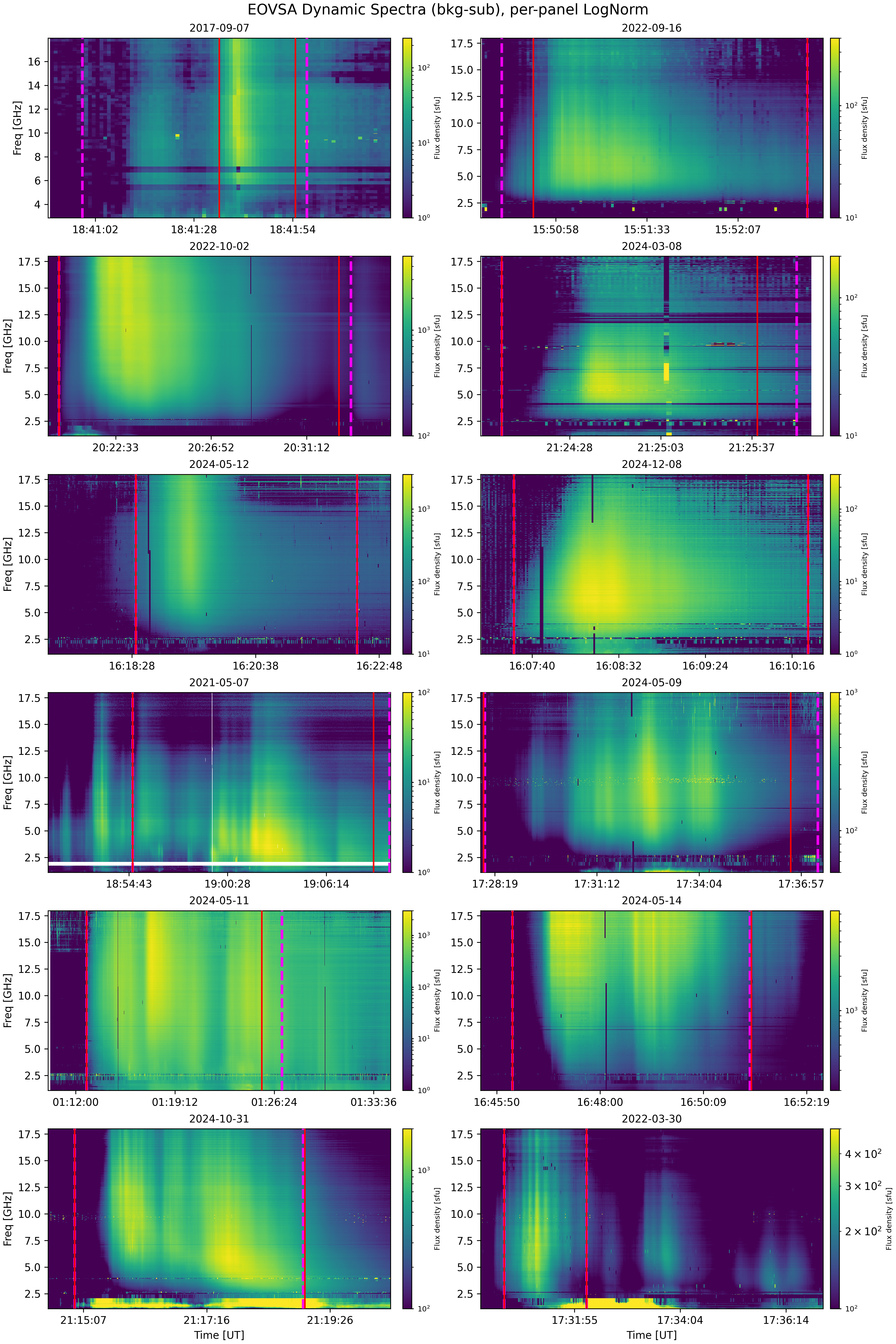}
\caption{ Gallery of EOVSA dynamic spectra (background-subtracted) of solar flares used in the analysis. The color scale shows flux density (sfu) in logarithmic normalization. Magenta dashed vertical lines mark the fitting interval, while red solid vertical lines indicate the time range displayed in the corresponding evolution panels.
\label{Fig:gallery_dyn_spectra}
}
\end{figure*}

Our flare selection, with the dynamics spectra shown in Figure\,\ref{Fig:gallery_dyn_spectra}, contains 12 impulsive flares of various GOES classes, from C4.5 to X8.8; however, the set is biased towards more powerful flares of the X class (7 events). This is motivated by the fact that radio bursts with stronger flux have, on average, larger area and, thus, they are better spatially resolved by the radio interferometer, which permits a more meaningful spectral fitting. In addition, when the radio flux is large, the resulting flare spectrum is less sensitive to subtraction of the preflare background than in the case of a weak flare. Yet, our list includes one C class and four M class flares.

The selected flares display a range of durations listed in Table\,\ref{table_combined_corr}. The shortest one lasted less than a half of a minute; the longest---10-15 minutes, while others from three to five minutes. A majority (7) of the bursts have only a single peak in the light curve permitting quantitative analysis; others display two or more peaks; one of them shows quasi-periodic pulsations (QPPs). The absolute peak fluxes cover a broad range, from about 100\,sfu to more than 7,000\,sfu, see Table\,\ref{table_combined_corr}.

%









\section{Spectral model fitting}

A link between the radio spectrum slope and the energy spectrum of highly relativistic electrons was understood in the early 1950s following the discovery of radio galaxies and other cosmic sources with unusual nonthermal spectra. 
A unique relationship between the slope of the radio spectrum $\gamma$ and the slope of the energy spectrum of relativistic electrons $\delta$, $\gamma=(\delta-1)/2$, was established when the theory of synchrotron radiation from power-law electron distributions was developed \citep[see, e.g., the review by][]{1969ARA&A...7..375G}. 

It was, therefore, tempting to establish a similar relationship for solar flares, where weakly and moderately relativistic electrons are a lot more numerous than the highly relativistic ones.
However, although there is a general trend that softer electron spectra produce steeper \mw\ spectra in the optically thin region, no  unique relationship between those indices exists in the case of the \gs\ emission produced by mildly relativistic electrons \citep[see, e.g.,][]{Fl_Meln_2003b}, where the \mw\ spectrum often contains both optically thin and thick regions separated by the spectral peak. We have already mentioned in the Introduction that the spectral index $\delta$ affects not only the optically thin but also optically thick part of the spectrum; thus, reliable inference of the spectral index requires analysis of the full spectrum that includes both optically thin and thick domains.  

The importance of the optically thick spectral region can be clarified with the following consideration. The optically thick \mw\ flux $S$ at a given frequency $f$ is defined solely by the product of the source area $A$ and the brightness temperature $T_b$ \citep[see, e.g., Eqn (2) in][]{2018ApJ...867...81F}. The source area is fixed to the pixel size, while the brightness temperature is defined by the effective temperature of nonthermal electrons contributing to the \gs\ opacity at the given frequency. It is clear that for progressively harder spectra (smaller $\delta$) the contribution of higher energy electrons becomes more and more important resulting in a progressively larger brightness temperature of the \gs\ emission. Therefore, the most robust diagnostics of the spectral index $\delta$ is possible when the entire \mw\ spectrum---comprising both the optically thin and thick regions---is sampled, which is the case of most data in our flare list. However, even in the cases when only the optically thick part of the spectrum is available, the model spectral fitting still yields useful information on the electron energy slope $\delta$.

\subsection{\gf Spectral fitting settings}
\label{S_fit_overview}

The core element of our analysis is the \mw\ spectrum obtained for a given time from a given location (pixel) on the \mw\ map. This spectrum with estimated uncertainties is forward-fitted by a cost source function, which takes into account both \gs\ emission from the nonthermal electrons and free-free emission from the ambient plasma described in the form of fast codes \citep[specifically, their fastest, continuous version described in][]{Fl_Kuzn_2010}. When data permitted, we used a source function that included a single uniform source with the area prescribed by the pixel size ($2\arcsec\times2\arcsec$) and an assumed depth (typically, equivalent to $8\arcsec$). This source is characterized by the following physical parameters: the magnetic field strength $B$; the viewing angle $\vartheta$, the thermal plasma number density $n_{th}$ and temperature $T$; and parameters of a power-law distribution of the nonthermal electrons, including the minimum and maximum energies, $E_{\min}$ and $E_{\max}$, the number density $n_{nth}$, and the spectral index $\delta$. In most of the considered cases the spectra are not (or only weakly) sensitive to $T$, $E_{\min}$, and $E_{\max}$; so we typically fixed them at the default values of $T=20$\,MK, $E_{\min}=15$\,keV, and $E_{\max}=5$\,MeV. The remaining five parameters--$B$, $\vartheta$, $n_{th}$, $n_{nth}$, and $\delta$--were kept free and were determined from the spectral fit. 

\begin{figure*}[!h]
\centering
\includegraphics[width=0.32\linewidth]{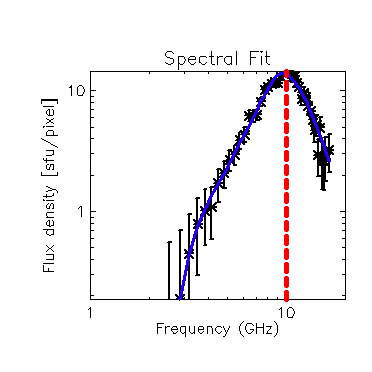}
\includegraphics[width=0.32\linewidth]{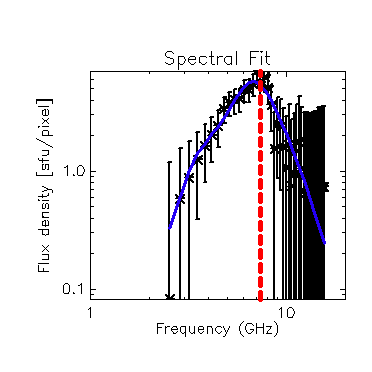}
\includegraphics[width=0.32\linewidth]{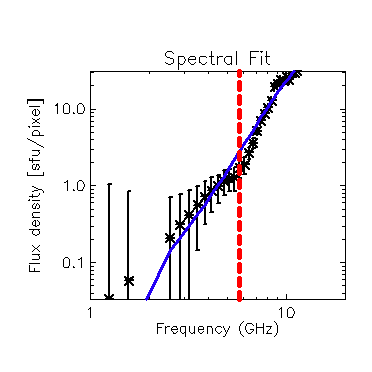}
\includegraphics[width=0.32\linewidth]{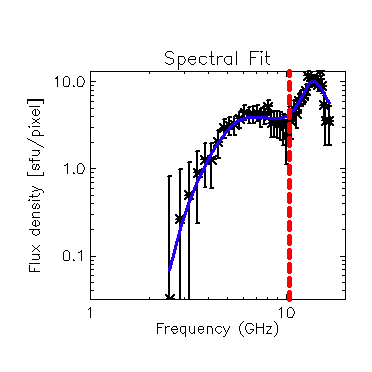}
\includegraphics[width=0.32\linewidth]{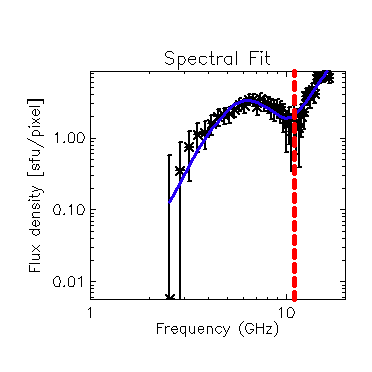}
\includegraphics[width=0.32\linewidth]
{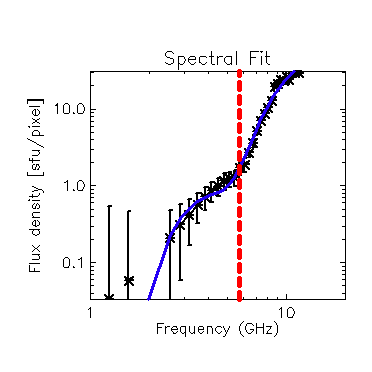}
\caption{Examples of spatially resolved \mw\ spectra and their corresponding fits. The first and second columns show spectra for the 2022-Oct-02 flare, while the last column for the 2024-May-14 flare. The first row gives examples of the fits with a single source model, the second row gives examples of two-component spectra and fits. Division onto single- and two-component cases is evident for the presented 2022-Oct-02 spectra that clearly show either one or two distinct spectral peaks. The last column offers a less evident case of a single-peak spectrum (the same in both panels), which, however, cannot be properly fit by a single-component model because of the break-up to the right of the red vertical line. Such breaks are indications of the presence of an additional spectral component that is taken into account in the bottom panel showing excellent fit to the data.
\label{Fig:spec_fit_examples}
}
\end{figure*}

There are cases where individual spatially resolved spectra deviate from those expected for a single uniform source. In such cases, we employed a nonuniform source model consisting of two uniform sources placed sequentially along the line of sight, and took into account the radiation transfer through the source located closer to the observer. Generally, this approach doubles the number of free parameters; however, when permitted by data, we assumed that the spectral indices of both sources are identical, which reduces the number of free parameters to nine. Figure\,\ref{Fig:spec_fit_examples} shows a representative set of single- and two-component spectra and their corresponding fits.

This spectral model fitting is performed using a dedicated tool, GSFIT, introduced in \citet{2020Sci...367..278F} and further upgraded since then to serve the user needs in its flexibility and available options. In particular, it permits a sequential spectral fitting over a selected region of interest (ROI) and time range. A smart search algorithm that combines a simplex minimization with several shakings of the solution \citep{Fl_etal_2009,Gary_etal_2013,2026ApJ...999..179F} is designed to find a true solution (a global minimum of the functional) and thus, the spectral fitting outcome should not depend on the initial trial values of the fit parameters. This is often the case, but there are situations when the outcome is not fully unique and depends on these trial parameters. To deal with such cases, GSFIT also permits refitting a subset of pixels with selected values of reduced $\chi^2$ metrics with a new set of initial parameters.

We performed this bulk spectral fitting for a selected set of solar flares and derived evolving maps of the mentioned physical parameters. Here, however, we focus on only one of them---the energy spectral index $\delta$---to recover typical patterns of spectral evolution of mildly relativistic electrons responsible for the \mw\ emission from solar flares.

\subsection{\gf Robustness of the spectral model fitting}

{\gf
The outcome parameters of the spectral fitting can, in principle, be correlated with each other or show degeneracy. A robust way to investigate these is to explore the full parameter space via, e.g., Monte Carlo Markov Chain (MCMC) simulations. MCMC simulations are very time-consuming and, thus, impractical in our case of massive spectral fitting. MCMC tests were performed in several instances to confirm the results of the GSFIT, which is better suited for the bulk fitting because it works much faster.

\citet{2020NatAs...4.1140C} reported MCMC tests for two spatially resolved spectra for a rising phase of the 2017-Sep-10 flare. The first of them contains a very restricted portion of the optically thin part of the \gs\ spectrum. The tests revealed correlations between $\delta$ and other fit parameters. Yet, the spectral index is well constrained with the median value of about 3.5 and a standard deviation of around 0.5. The second reported spectrum has two components with an extended optically thin portion. In this case, spectral indices associated with both components are well constrained with accuracy better than 10\%; they display only modest correlations with other parameters.

\citet{2022Natur.606..674F} performed MCMC simulations for an entire single time frame, 15:58 UT, of the EOVSA \mw\ data of the main phase of the 2017-Sep-10 flare. The MCMC parameter maps were found consistent with the GSFIT-provided parameter maps, which validates the GSFIT approach to the bulk spectral fitting. \citet{2022Natur.606..674F} also reported results of the MCMC simulations for two different representative spatially resolved spectra---one with both optically thin and thick parts separated by a spectral peak and another one with only a rising (optically thick) portion of the spectrum. In the first case, the spectral index $\delta$ shows weak correlations with the inferred magnetic field $B$ and the number density of nonthermal electrons $n_{\rm nth}$. In the second case, only a correlation with the magnetic field is present. Indeed, the optically thick spectrum does not depend on the number density of nonthermal electrons, but only on the effective energy of electrons providing dominant contribution to the \gs\ opacity at a given frequency. In both cases the spectral index is well constrained being $\delta=5.8\pm0.3$ in the first case and $\delta=2.1\pm0.3$ in the second case. This conclusion holds for smooth spatially resolved spectra with high signal-to-noise ratio, such as illustrated in Extended Data Fig. 1 from \citep{2022Natur.606..674F}. This figure shows a variety of spatially resolved spectra with various shapes and signal-to-noise ratios of which only areas inscribing spectra with good signal-to-noise ratio are employed for quantitative analysis. Based on the described MCMC simulations, we conclude that the inference of the electron spectral index from the model spectral fitting of the spatially localized \mw\ spectra is robust provided the signal-to-noise ratio is high, which is is the case with our dataset.
}



\subsection{Spectral Index Evolution}

\begin{figure*}\centering
\includegraphics[width=0.98\linewidth]{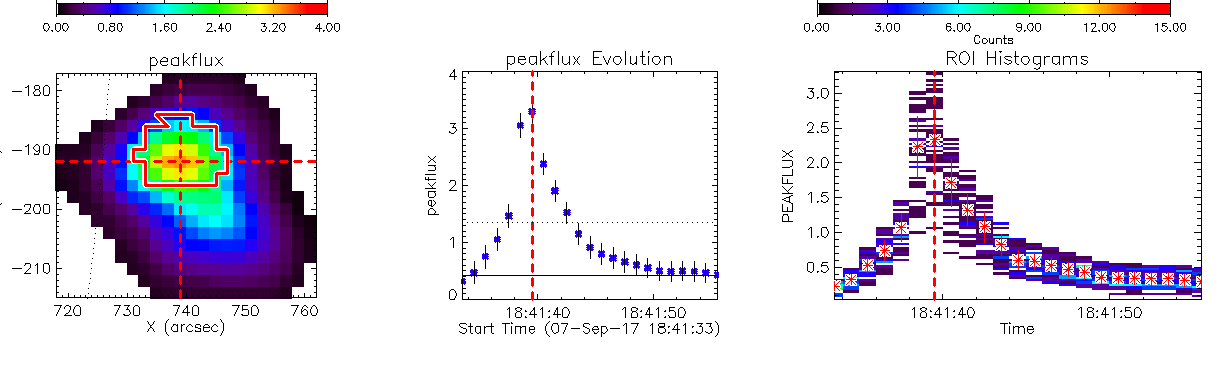}
\includegraphics[width=0.98\linewidth]{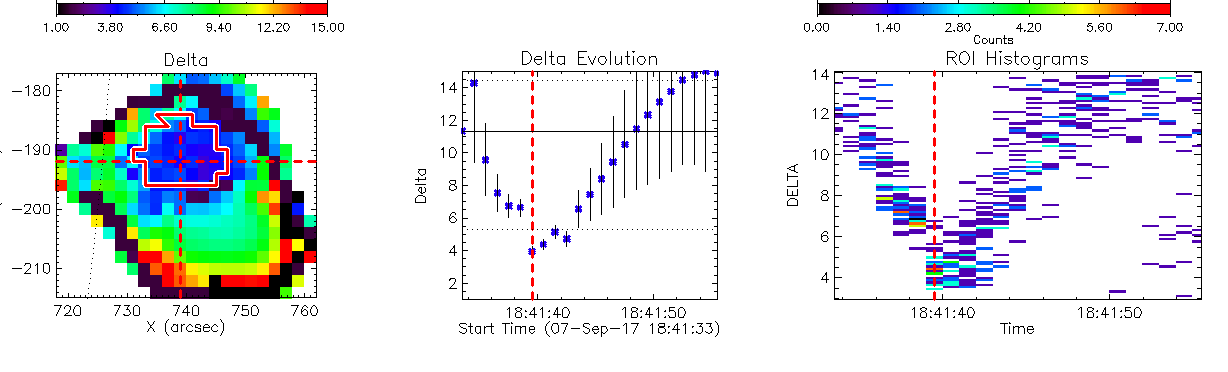}
\includegraphics[width=0.98\linewidth]{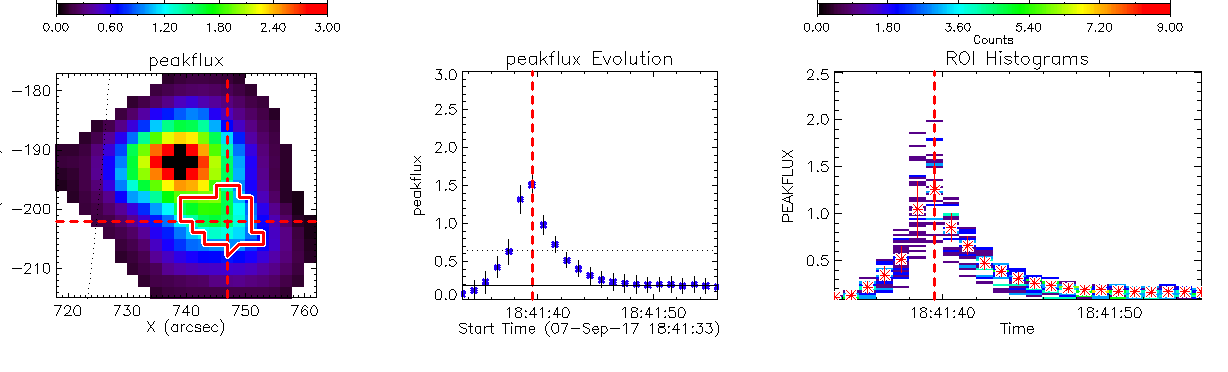}
\includegraphics[width=0.98\linewidth]{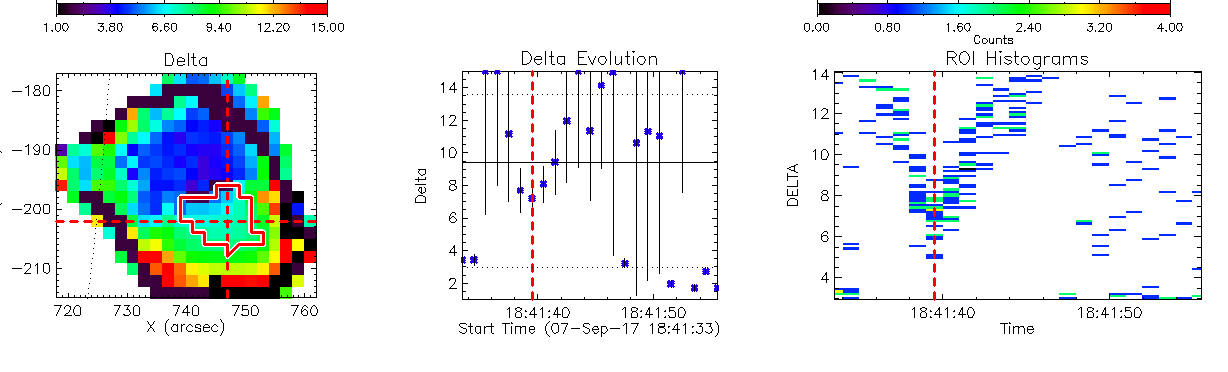}
\caption{Left panels show maps of the selected parameter, the peak flux or the spectral index, obtained for the 2017-Sep-07 flare in the peak flux time frame indicated by the red dashed vertical lines in the middle and right panels. Middle panels show evolution of the corresponding parameter in the pixel indicated by the red cursor in the maps on the left. Right panels show evolving 2D histograms of the corresponding parameter within the region of interest (ROI) outlined by the red polygon in the maps on the left. We use two different ROIs that correspond to two different flare regions (flaring loops) identified in \citet{2025ApJ...988..260F}. Red asterisks with the error bars show the corresponding median values with their 1$\sigma$ uncertainties. Both ROIs in this event reveal the same SHS trend of the spectrum.
\label{Fig:delta_2017-Sep-07}
}
\end{figure*}

\begin{figure*}\centering
\includegraphics[width=0.98\linewidth]{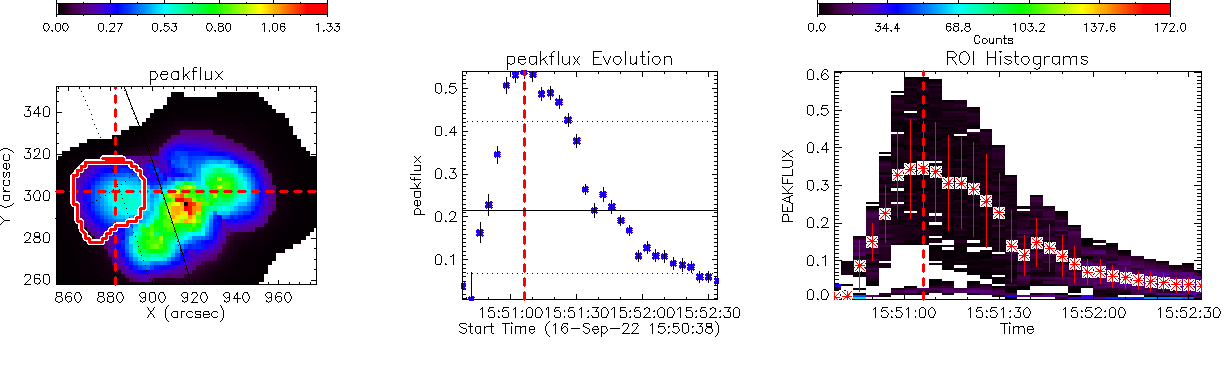}
\includegraphics[width=0.98\linewidth]{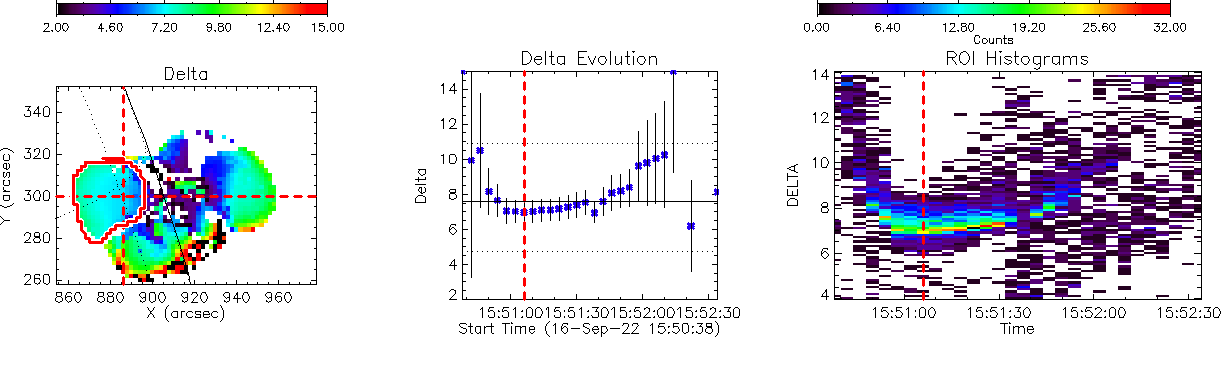}
\includegraphics[width=0.98\linewidth]{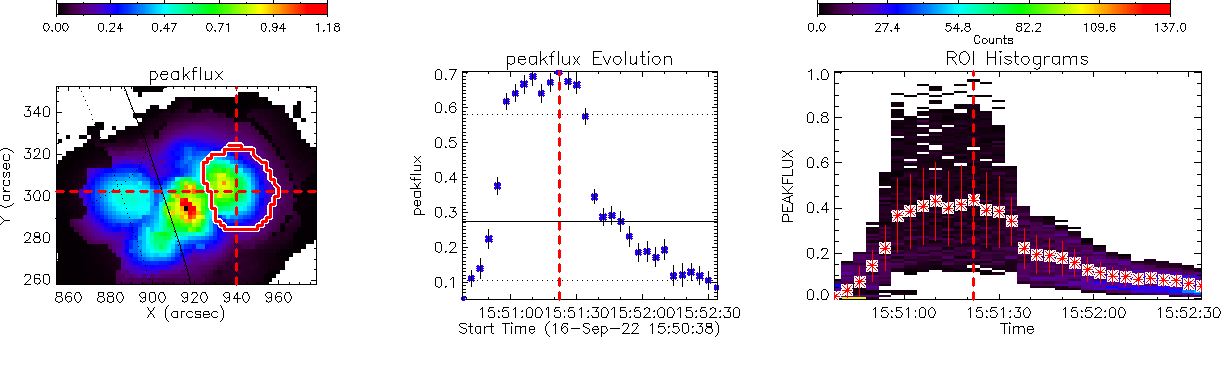}
\includegraphics[width=0.98\linewidth]{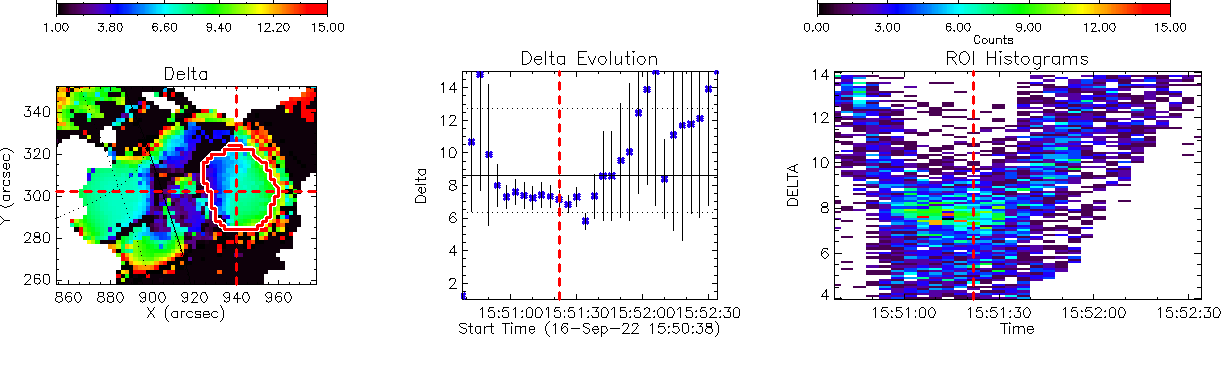}
\caption{Left panels show maps of the selected parameter, the peak flux or the spectral index, obtained for the 2022-Sep-16 flare in the peak flux time frame indicated by the red dashed vertical lines in the middle and right panels. Middle panels show evolution of the corresponding parameter in the pixel indicated by the red cursor in the maps on the left. Right panels show evolving 2D histograms of the corresponding parameter within the region of interest (ROI) outlined by the red polygon in the maps on the left. Two different ROIs are manually selected to inscribe two differently behaved regions of the flare. Red asterisks with the error bars show the corresponding median values with their 1$\sigma$ uncertainties. Both ROIs in this event reveal the same SHS trend of the spectrum.
\label{Fig:delta_2022-Sep-16}
}
\end{figure*}

\begin{figure*}\centering
\includegraphics[width=0.98\linewidth]{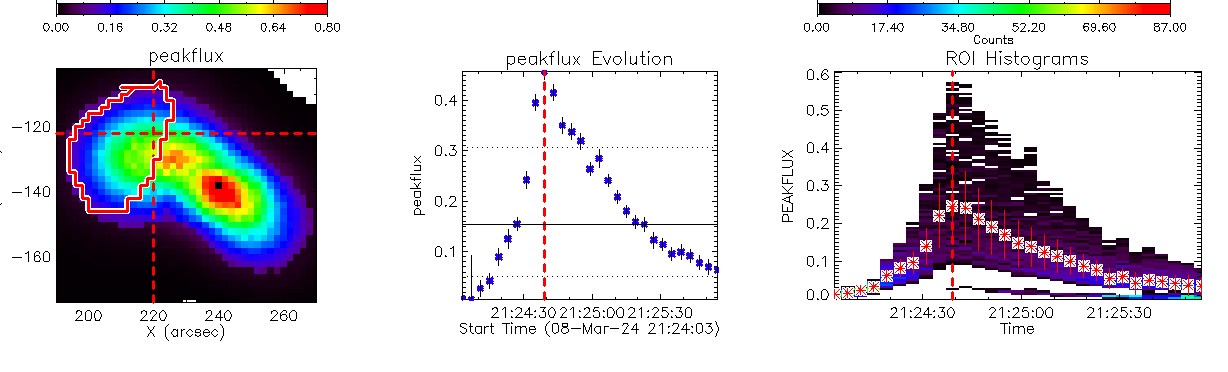}
\includegraphics[width=0.98\linewidth]{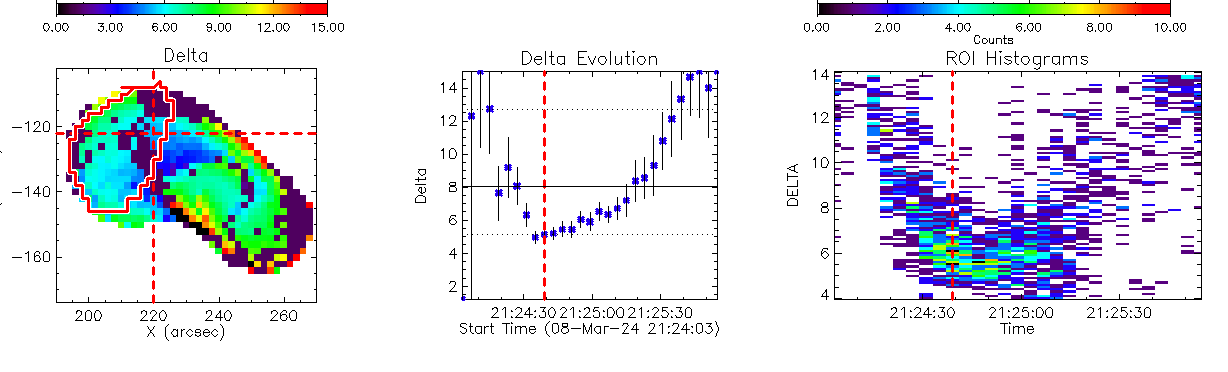}
\includegraphics[width=0.98\linewidth]{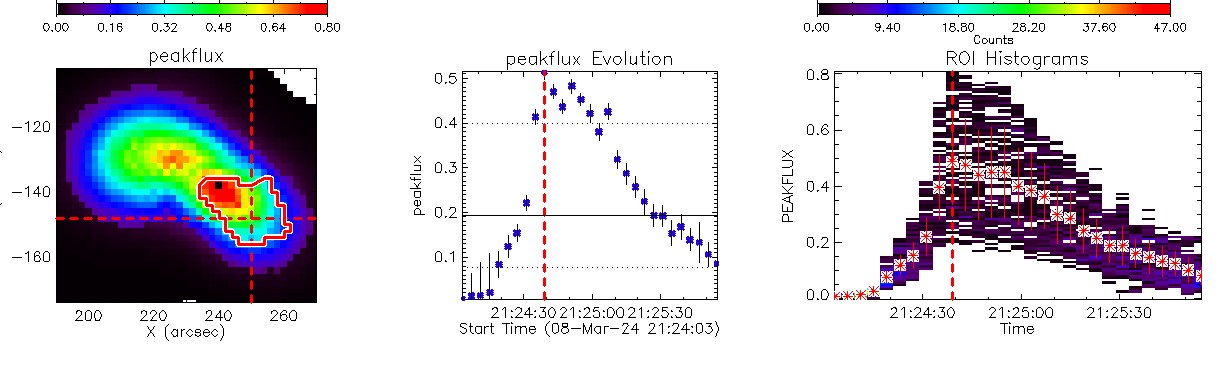}
\includegraphics[width=0.98\linewidth]{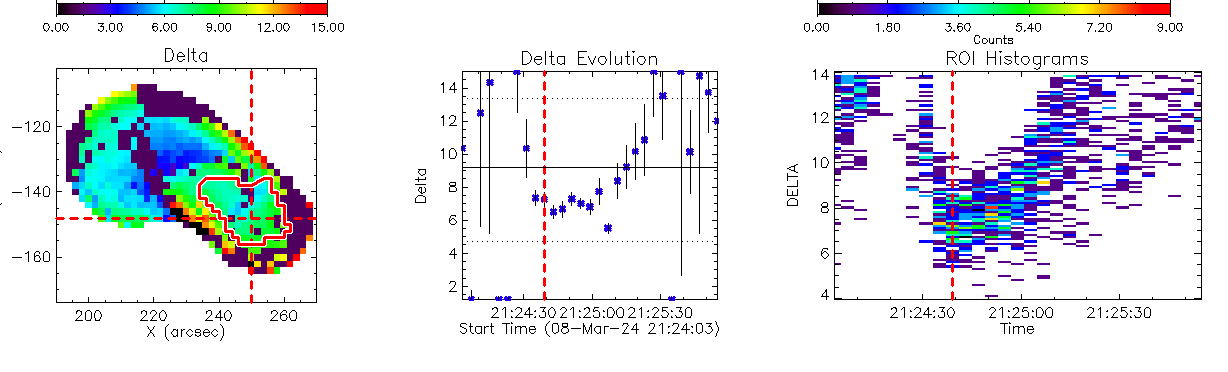}
\caption{Left panels show maps of the selected parameter, the peak flux or the spectral index, obtained for the 2024-Mar-08 flare in the peak flux time frame indicated by the red dashed vertical lines in the middle and right panels. Middle panels show evolution of the corresponding parameter in the pixel indicated by the red cursor in the maps on the left. Right panels show evolving 2D histograms of the corresponding parameter within the region of interest (ROI) outlined by the red polygon in the maps on the left. Two different ROIs are manually selected to inscribe two differently behaved regions of the flare. Red asterisks with the error bars show the corresponding median values with their 1$\sigma$ uncertainties. Both ROIs in this event reveal the same SHS trend of the spectrum.
\label{Fig:delta_2024-Mar-08}
}
\end{figure*}

\begin{figure*}\centering
\includegraphics[width=0.98\linewidth]{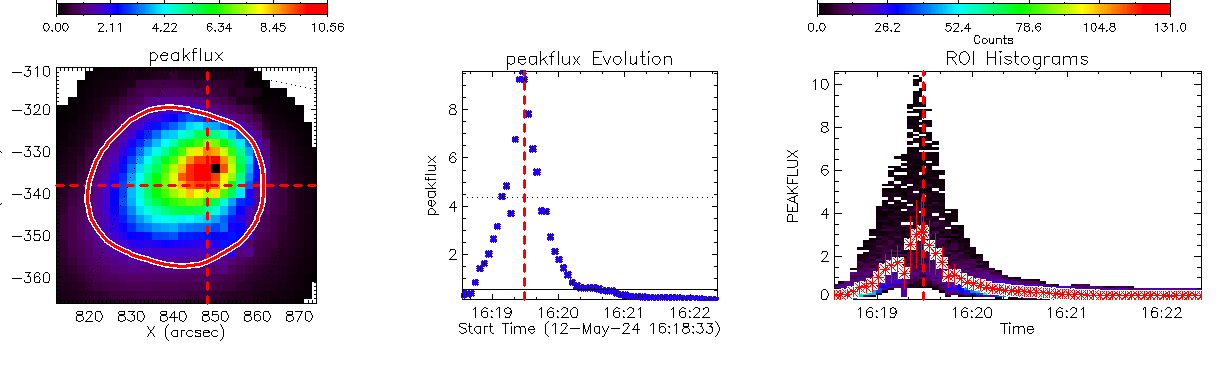}
\includegraphics[width=0.98\linewidth]{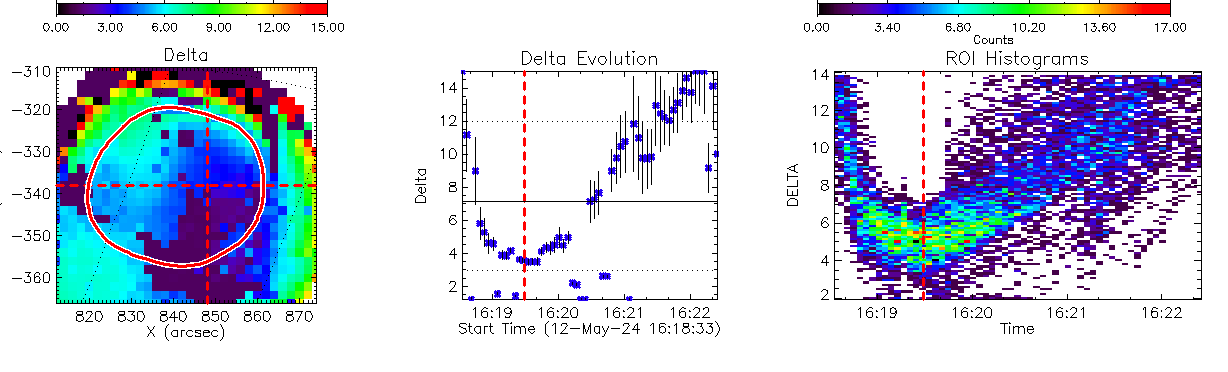}
\includegraphics[width=0.98\linewidth]{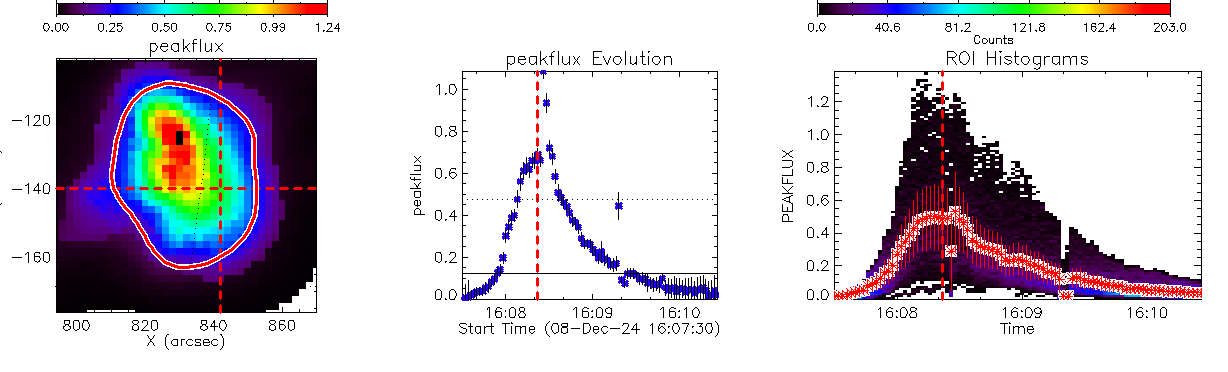}
\includegraphics[width=0.98\linewidth]{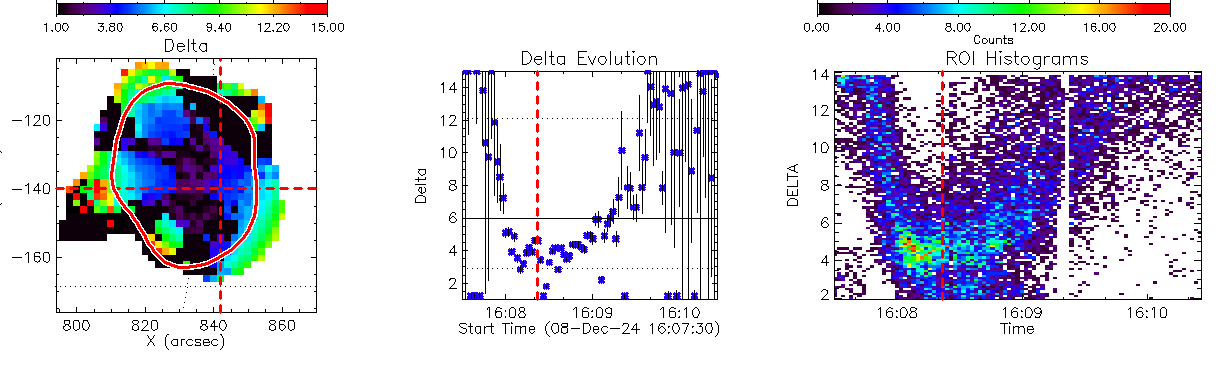}
\caption{Left panels show maps of the selected parameter, the peak flux and the spectral index, obtained for two flares in the peak flux time frame indicated by the red dashed vertical lines in the middle and right panels. Middle panels show evolution of the corresponding parameter in the pixel indicated by the red cursor in the maps on the left. Right panels show evolving 2D histograms of the corresponding parameter within the region of interest (ROI) outlined by the red polygon in the maps on the left. The ROIs are selected to inscribe most of the \mw\ sources as  15\% contours at the 8.71\,GHz map taken at 16:19:29\,UT for the 2024-May-12 flare and at the 6.44\,GHz map taken at 16:08:38\,UT for the 2024-Dec-08 flare. Red asterisks with the error bars show the corresponding median values with their 1$\sigma$ uncertainties. 
\label{Fig:delta_two_single_cases}
}
\end{figure*}

\begin{figure*}\centering
\includegraphics[width=0.98\linewidth]{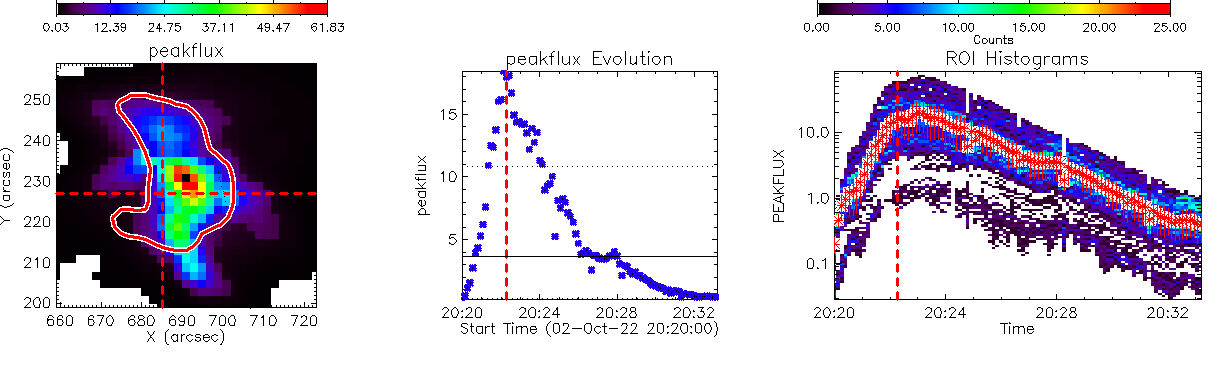}
\includegraphics[width=0.98\linewidth]{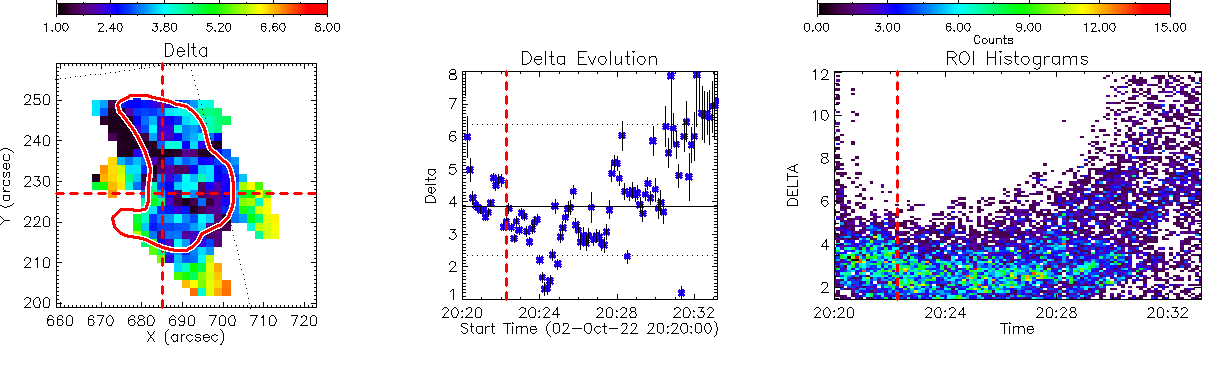}
\includegraphics[width=0.98\linewidth]{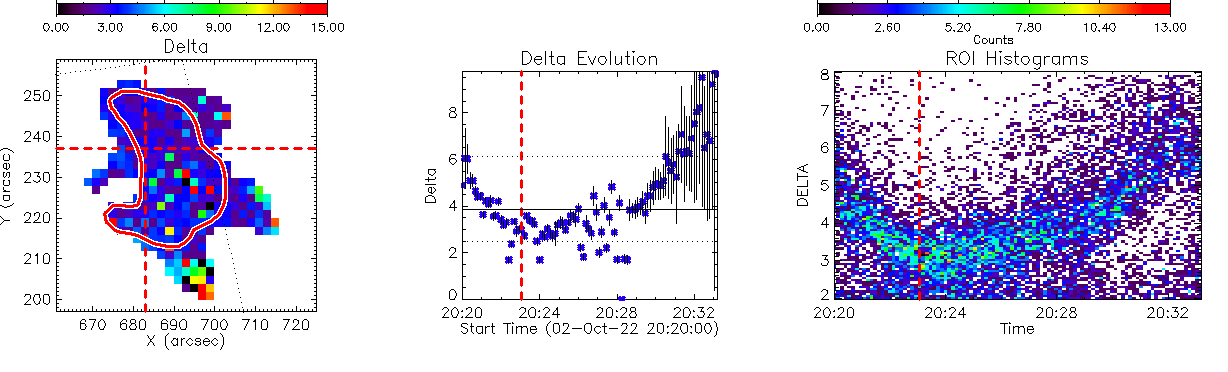}
\includegraphics[width=0.98\linewidth]{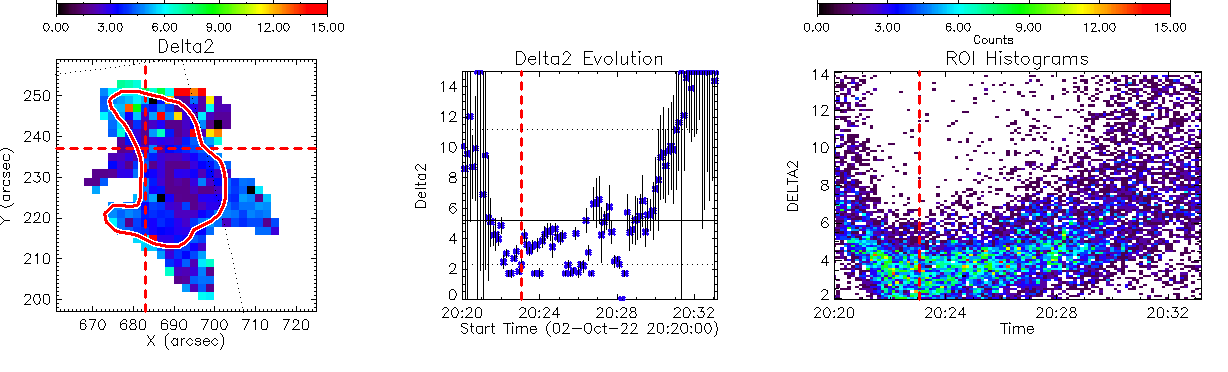}
\caption{Left panels show maps of the selected parameter, the peak flux and the spectral index, obtained for the 2022-Oct-02 flare in the peak flux time frame indicated by the red dashed vertical lines in the middle and right panels. Middle panels show evolution of the corresponding parameter in the pixel indicated by the red cursor in the maps on the left. Right panels show evolving 2D histograms of the corresponding parameter within the region of interest (ROI) outlined by the red polygon in the maps on the left. The ROI is selected to inscribe most of the \mw\ source as a 15\% contour at the 9.36\,GHz map taken at 20:23:04\,UT. Red asterisks with the error bars show the corresponding median values with their 1$\sigma$ uncertainties. The second row shows results of the spectral model fitting using a single source on the LOS, while the third and fourth show evolution of two spectral indices in the model with two sources; see text.
\label{Fig:delta_2022-Oct-02}
}
\end{figure*}

Figure\,\ref{Fig:delta_2017-Sep-07}--\ref{Fig:delta_2022-Oct-02} display the spectral fitting results for flares with only one peak in the light curves. Their common behavior is the prominent SHS spectral pattern during the rise-peak-decay evolution of the \mw\ flux. 
The hardening and softening rates differ from each other: the softening occurs slower and takes more time than the initial hardening, resulting in an event-specific asymmetry of the curves describing the spectral index evolution. Another common feature is a very broad range of the spectral index variation---between rather small values, $\delta=2-3$, and very high, $\delta=15$, which was adopted as the upper bound of $\delta$ while performing the spectral fitting; this implies that even larger values are possible \citep{2021ApJ...908L..55C}. This range of $\delta$ variation is much broader than that reported based on the HXR data analysis
\citep{1985SoPh..100..465D,2004A&A...426.1093G,2018ApJ...867...84G,2024A&A...684A.215C}.

\begin{figure*}\centering
\includegraphics[width=0.98\linewidth]{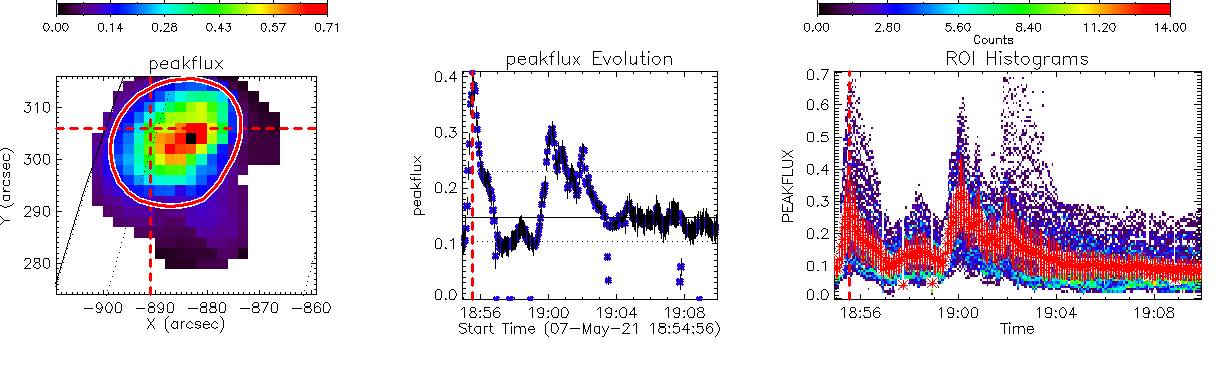}
\includegraphics[width=0.98\linewidth]{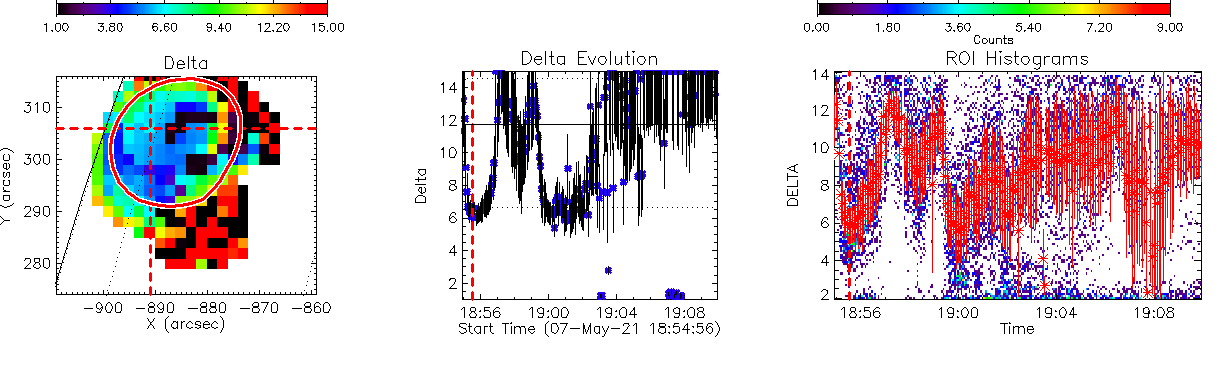}
\includegraphics[width=0.98\linewidth]{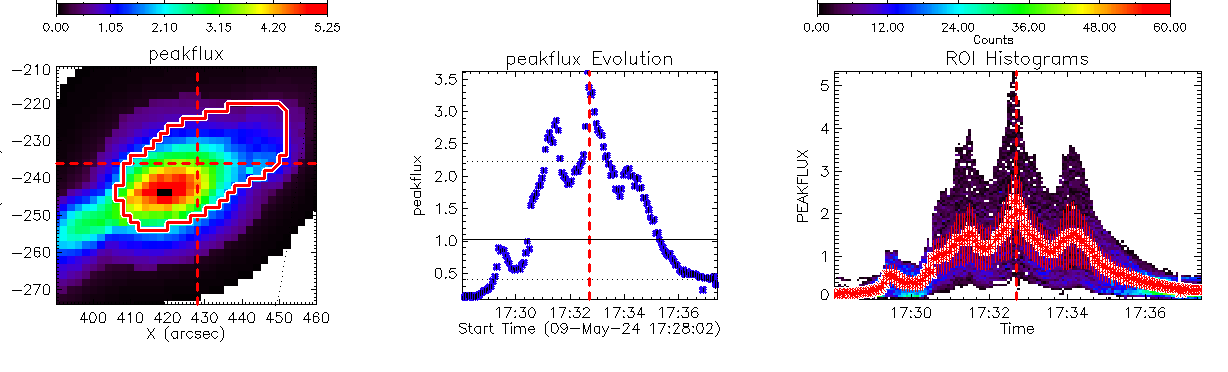}
\includegraphics[width=0.98\linewidth]{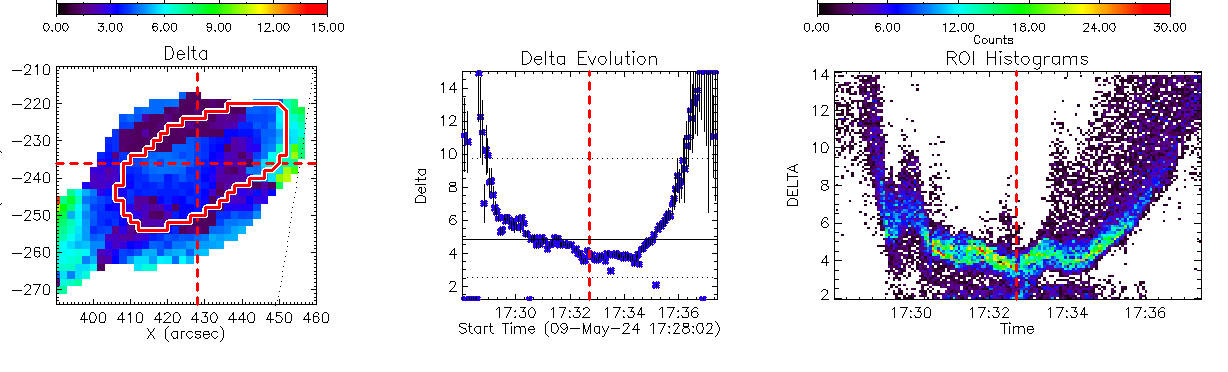}
\caption{Left panels show maps of the selected parameter, the peak flux and the spectral index, obtained for two multi-peak flares in the peak flux time frame indicated by the red dashed vertical lines in the middle and right panels. Middle panels show evolution of the corresponding parameter in the pixel indicated by the red cursor in the maps on the left. Right panels show evolving 2D histograms of the corresponding parameter within the region of interest (ROI) outlined by the red polygon in the maps on the left. The ROIs are selected to inscribe most of the \mw\ source as a 20\% contour at the 6.44\,GHz map taken at 18:55:28\,UT for the 2021-May-07 flare and  manually selected based on a $\chi^2$ map for the 2024-May-09 flare. Red asterisks with the error bars show the corresponding median values with their 1$\sigma$ uncertainties. 
\label{Fig:delta_two_multi_peak_cases}
}
\end{figure*}

\begin{figure*}\centering
\includegraphics[width=0.98\linewidth]{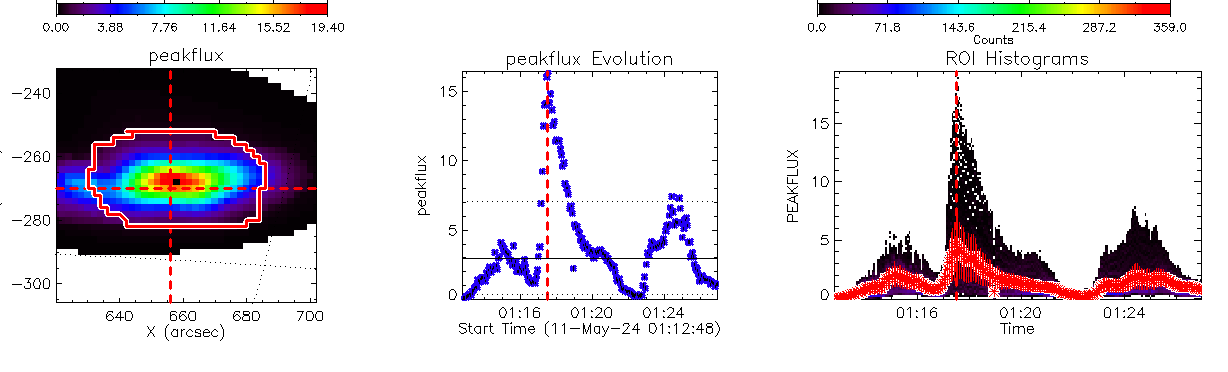}
\includegraphics[width=0.98\linewidth]{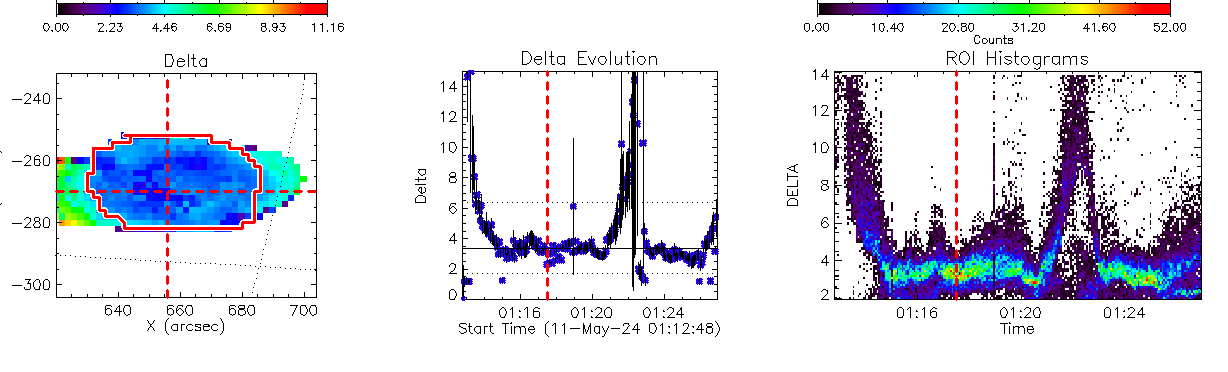}
\includegraphics[width=0.98\linewidth]{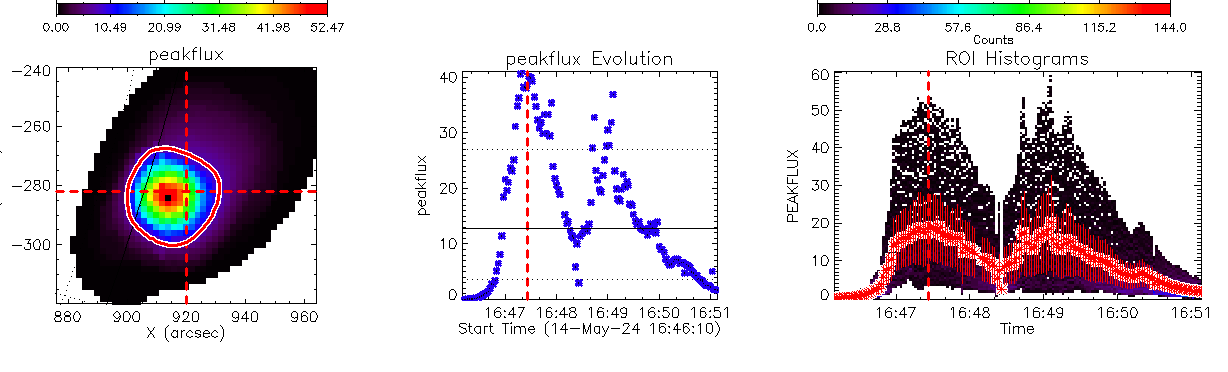}
\includegraphics[width=0.98\linewidth]{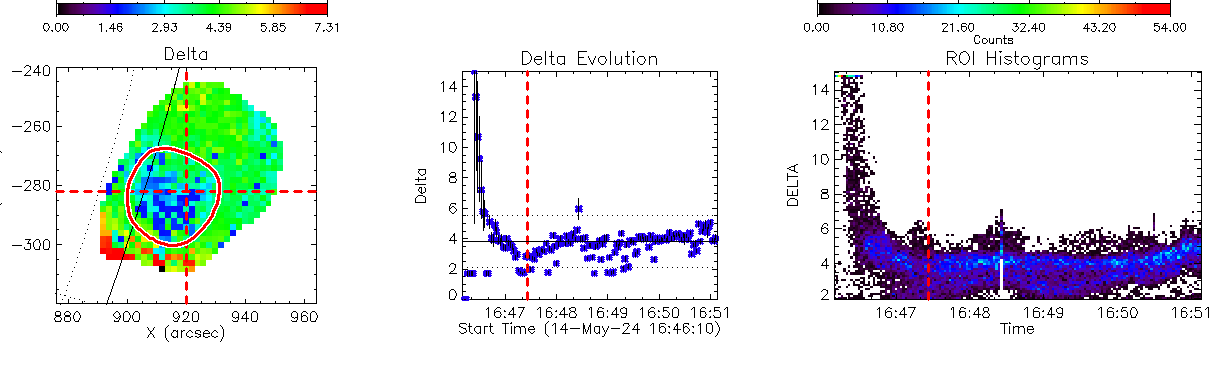}
\caption{Left panels show maps of the selected parameter, the peak flux and the spectral index, obtained for two multi-peak flares in the peak flux time frame indicated by the red dashed vertical lines in the middle and right panels. Middle panels show evolution of the corresponding parameter in the pixel indicated by the red cursor in the maps on the left. Right panels show evolving 2D histograms of the corresponding parameter within the region of interest (ROI) outlined by the red polygon in the maps on the left. The ROI is manually selected based on the $\delta$ map for the 2024-May-11 flare
and to inscribe most of the \mw\ source as a 15\% contour at the 10.66\,GHz map taken at 16:47:26\,UT for the 2024-May-14 flare. Red asterisks with the error bars show the corresponding median values with their 1$\sigma$ uncertainties. 
\label{Fig:delta_two_multi_peak_cases_SH}
}
\end{figure*}

\begin{figure*}\centering
\includegraphics[width=0.98\linewidth]{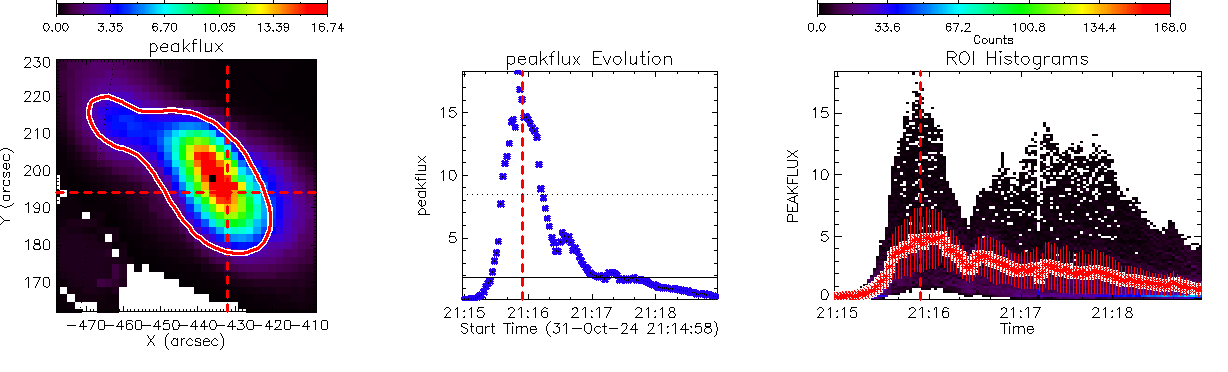}
\includegraphics[width=0.98\linewidth]{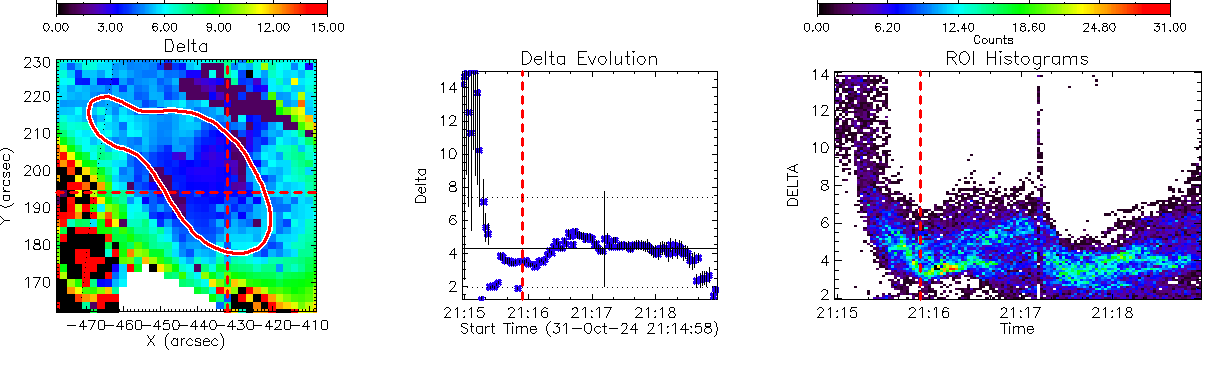}
\caption{Left panels show maps of the selected parameter, the peak flux and the spectral index, obtained for a multi-peak flare in the peak flux time frame indicated by the red dashed vertical lines in the middle and right panels. Middle panels show evolution of the corresponding parameter in the pixel indicated by the red cursor in the maps on the left. Right panels show evolving 2D histograms of the corresponding parameter within the region of interest (ROI) outlined by the red polygon in the maps on the left. The ROI is selected to inscribe most of the \mw\ source as a 15\% contour at the 10.01\,GHz map taken at 21:15:54\,UT. Red asterisks with the error bars show the corresponding median values with their 1$\sigma$ uncertainties. 
\label{Fig:delta_2024-10-31_SH}
}
\end{figure*}

\begin{figure*}\centering
\includegraphics[width=0.98\linewidth]{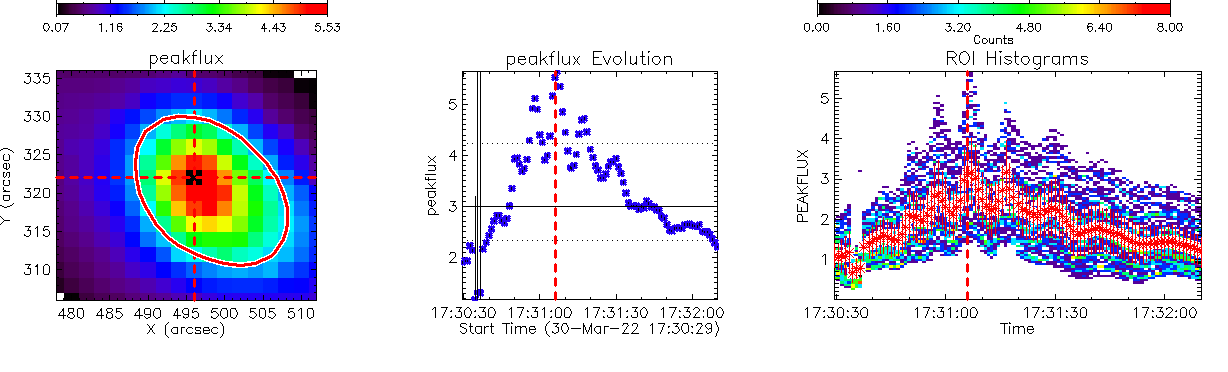}
\includegraphics[width=0.98\linewidth]{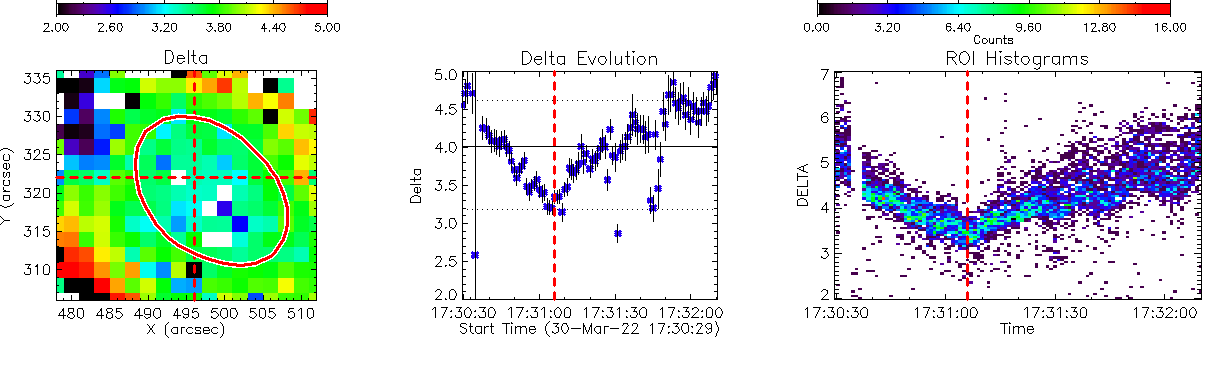}
\includegraphics[width=0.98\linewidth]{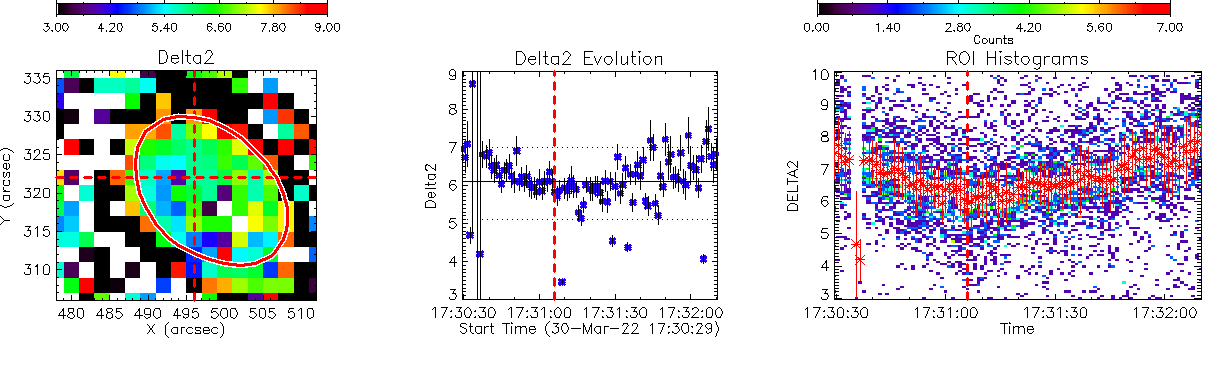}
\caption{Left panels show maps of the selected parameter, the peak flux and the spectral index, obtained for a multi-peak flare in the peak flux time frame indicated by the red dashed vertical lines in the middle and right panels. Middle panels show evolution of the corresponding parameter in the pixel indicated by the red cursor in the maps on the left. Right panels show evolving 2D histograms of the corresponding parameter within the region of interest (ROI) outlined by the red polygon in the maps on the left. The ROI is selected to inscribe the core of the \mw\ source with most pronounced QPPs as a 30\% contour at the 13.26\,GHz map taken at 17:31:07\,UT. Red asterisks with the error bars show the corresponding median values with their 1$\sigma$ uncertainties. 
\label{Fig:delta_2022-03-30}
}
\end{figure*}

The details are different both between different flares and between different locations of the same flare. For example, in the 2017-Sep-07 flare that contained two distinct flaring loops \citep{2025ApJ...988..260F} at the upper left and the bottom right of the image, in the upper ROI $\delta$ varied between 14-4-15, while in the lower ROI---only between 15-5-15; thus, the lower region displayed overall softer spectrum. The hardening and softening rates are similar to each other in this case; although the softening occurs slightly slower and takes more time than the initial hardening like in other events. 

The 2022-Sep-16 and 2024-Mar-08 flares 
display a similar trend but lasting four-five times longer, about two minutes each. 
Figures~\ref{Fig:delta_2022-Sep-16} and \ref{Fig:delta_2024-Mar-08} analyze two distinct ROIs each separately because the spatially resolved spectra are consistent with the employed uniform source model inside those ROIs. The areas between those ROIs show evidence of source nonuniformity (like a double peak spectral structure), which we discarded from the analysis of these flares. The range of $\delta$ variation is 15-(4-6)-15 during the burst. Two other single-peak flares shown in Fig.\,\ref{Fig:delta_two_single_cases} are fully consistent with the earlier described examples. 

A similar trend, but lasting about 12 minutes within a narrower range of $\delta$ between $\sim2-3$ and $\sim6-8$, is detected in the 2022-Oct-02 flare. Originally, we fitted this event with a single uniform source model, whose outcome is displayed in the second row of Fig.\,\ref{Fig:delta_2022-Oct-02}, but then we realized that the spectra show evidence of nonuniformity in many instances \citep{2026ApJ...999..179F}. We then repeated the spectral fitting assuming a two-source model and two independent $\delta$; the outcome of this fitting run is displayed in the third and fourth rows. One may see that the conclusion of a prominent SHS spectral evolution is robust and does not depend on the selected fitting model. 

The remaining cases contain flares with more than one peak on the light curves. The generally revealed SHS spectral trend is clearly seen in most if not all individual peaks. The $\delta$ variation is very strong during some of them, while more modest during others. For example, in the 2021-May-07 event, Fig.\,\ref{Fig:delta_two_multi_peak_cases}, $\delta$ varies between 6 and 15 up and down several times. On the contrary, in the 2024-May-09 flare shown in the same Figure, $\delta$ is large at the initial rise and final decay phases, but varies between $\sim3$ and $\sim6$ during all intermediate peaks. Nevertheless, the spectral index evolution is in anticorrelation with the peak flux, thus following the SHS evolutionary pattern.

The 2024-May-11 flare, Fig.\,\ref{Fig:delta_two_multi_peak_cases_SH}, shows a similar apparent trend; however, the spectrum remains hard in the decay phases or even hardens unless a new emission episode has started to grow. Here, we perhaps deal with a combination of SHS and SH(H) evolutionary patterns. The 2024-May-14 flare shows a similar trend with only a modest softening during the very late decay phase. The 2024-Oct-31, Fig.\,\ref{Fig:delta_2024-10-31_SH}, also shows rather hard spectra in the decay phase and, perhaps, even some hardening towards the end of its evolution. 

Finally, the 2022-Mar-30 clearly displays a train of QPPs with period of 7--10\,s; Fig.\,\ref{Fig:delta_2022-03-30}. The shape of spatially resolved spectra requires a fitting model with two distinct sources with different spectral indices, $\delta$ and $\delta_2$. Both show SHS evolutionary patterns throughout the duration of the analyzed \mw\ burst and also during each QPP episode. We conclude that the pronounced SHS pattern is an inherent property of the electron acceleration in solar flares.




\section{Discussion}


\begin{figure*}
\includegraphics[width=0.49\linewidth]{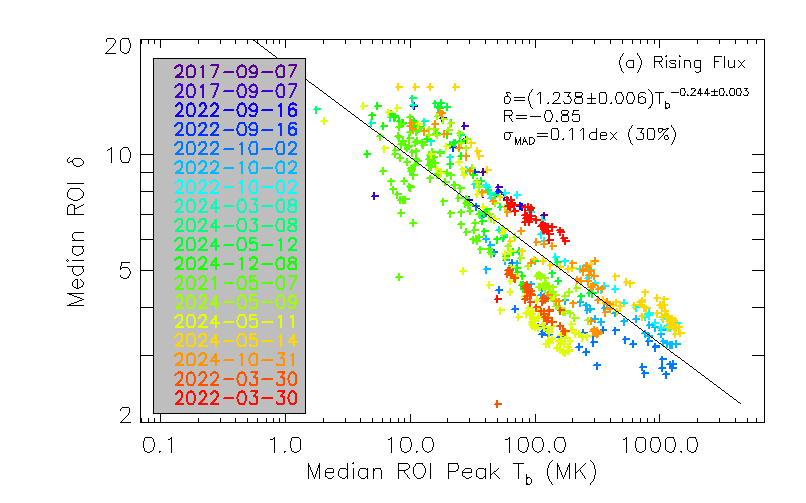}
\includegraphics[width=0.49\linewidth]{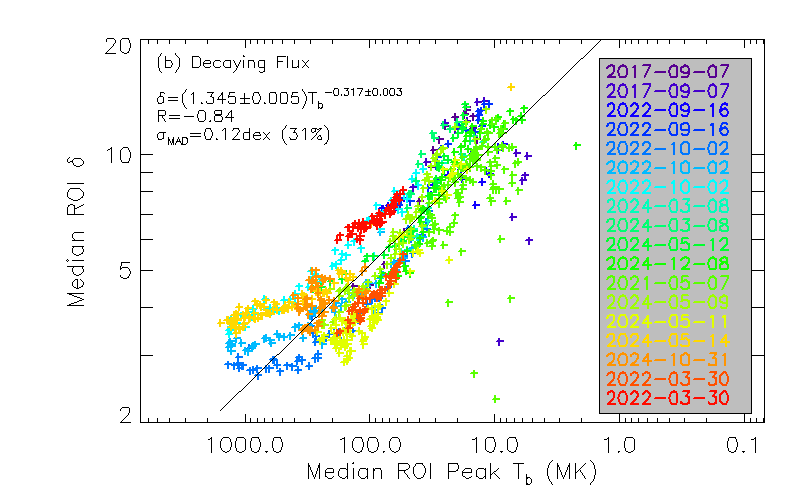}
\caption{\gf Spectral slope $\delta$ versus brightness temperature at the spectral-peak flux for the rising (a) and decaying (b) phases, color-coded by events as indicated in the legends. Each point represents the median peak brightness temperature and median $\delta$ within a given ROI for a single time bin. In panel~(b), the $x$-axis is inverted to emulate the temporal evolution. Note a flat region in panel (b) at large $T_B$ end of the plot, which is not present in panel (a). The solid lines show least-square power-law fits, with the fit parameters, their uncertainties, the linear correlation coefficient $R$, and the robust MAD-based scatter estimate $\sigma_{\mathrm{MAD}}$ listed in the legends. \label{Fig:HID}}
\end{figure*}

We have reported spectral evolution of nonthermal mildly relativistic electrons accelerated in a subset of impulsive solar flares inferred from the model spectral fitting of the \mw\ imaging spectroscopy data obtained with EOVSA. The main findings of our analysis are as follows. In a vast majority of cases (all but one) the population of nonthermal electrons first appears with a very steep spectrum with the index $\delta=10-15$. In one case of the 2022-10-02 flare, the initial spectral indices are within the range $\delta=5-10$. In all cases, this spectrum then hardens rather quickly, reaching values typically of $\delta=3-5$, while occasionally up to 6 or down to 2. This fast hardening is in full agreement with a similar finding of \citet{2021ApJ...908L..55C} obtained from a joint fit of HXR and \mw\ data for the rising phase of the eruptive long-duration 2017-Sep-10 flare.

While the rise phases show a unique trend in all the considered cases, the decay phases show somewhat more diversity. Nevertheless, all flares with a single temporal peak show a prominent softening in the decay phase such as the spectral index typically increases to roughly same values as in the very beginning of the burst. Moreover, this SHS pattern takes place during most of the individual peaks in the considered flares with more than one temporal peak. For some peaks, however, there are indications of continuing hardening during their decay phase, which then gives a way to a soft spectrum associated with the rise phase of the next temporal peak. In rare cases, the spectral softening at the final decay phase is rather modest or non-existent. 

{\gf Figure\,\ref{Fig:HID} offers a global view on this spectral evolution by showing a hardness-intensity diagram. 
We plot median values of ROIs' peak flux expressed here in terms of the brightness temperature, $T_B\propto S_{\rm peak}/f_{\rm peak}^2$, and spectral index for all time bins color-coded for all events separately for the impulsive (panel a) and decay (panel b) phases. 
Solid lines indicate linear fits to the log-linear data, with the corresponding Pearson correlation coefficients ($R$) displayed in the legends.
Both plots show a well-defined trend: higher brightness temperatures are systematically associated with harder electron spectra, i.e., smaller $\delta$. 
To quantify the dispersion around the trend lines robustly against outliers, the legends also include the Median Absolute Deviation (MAD) scatter estimate,
$\sigma_{\rm MAD} = 1.4826 \times \operatorname{median}\left( |\Delta\delta_i - \operatorname{median}(\Delta\delta)| \right)$,
where $\Delta\delta_i$ represents the residuals of the spectral index from the linear fit.
Similar plots that employ the peak flux rather than the brightness temperature show the same trends but with somewhat larger scatter, which is expected based on the discussion in Section\,\ref{S_fit_overview}.}


The $T_B$--$\delta$ relation is notable because the events in the sample differ substantially by their location on the disk, duration, flux level, spectral peak frequency, morphology, and fitted source structure; yet the event-resolved point clouds follow 
similar slopes in the $T_B$--$\delta$ plane. Thus, the flares do not merely show the same qualitative sense of evolution, with brighter states being harder; they appear to share a similar hardness--brightness scaling. This suggests that $T_B$ and $\delta$ are not primarily tracing the global scale or the lifetime of a flare, but rather the instantaneous condition of the microwave-emitting electron population. In this view, different flares may spend different amounts of time in the acceleration phase, but while active they follow a common spectral-brightness trend: high-brightness states are systematically associated with harder electron spectra, and low-brightness states with softer spectra.

At a qualitative level, this trend is physically natural: more intense energy release should produce a larger population of high-energy electrons and therefore brighter \mw\ emission with a harder spectrum. The nontrivial point is that this behavior appears as a reproducible $T_B$--$\delta$ relation across events with different durations, morphologies, and flux levels. This suggests that the relation is not controlled mainly by the global size or duration of each flare, but by a fundamental local acceleration condition that is similar across events. In other words, different flares may have different total energy budgets and lifetimes, but their \mw-emitting electrons appear to pass through similar instantaneous acceleration states.

This coupling is consistent with acceleration scenarios in which the spectral index changes with the instantaneous acceleration conditions, rather than remaining fixed while only the electron number changes. In stochastic or reconnection-driven turbulent acceleration, for example, stronger turbulence, faster acceleration, or improved confinement can simultaneously increase the number of mildly relativistic electrons and harden their spectrum \citep[e.g.,][]{1979ApJ...227.1072B,2004A&A...426.1093G,2006A&A...458..641G,Byk_Fl_2009}. A cross-event trend may then arise if both $T_B$ and $\delta$ are governed primarily by a common dimensionless control parameter, such as the ratio of acceleration rate to escape/loss rate \citep[see, e.g., Chapter 11 in][]{FT_2013}, rather than by the absolute duration or total energy of the flare. This ratio is physically important because it determines how far electrons can be driven up the energy distribution before they leave the acceleration region or lose energy. When acceleration is fast compared with escape and losses, electrons can populate higher energies more efficiently \citep{FT_2013}, producing both a harder spectrum and larger \mw\ brightness. When escape or losses dominate, the high-energy tail is depleted or weakly replenished, giving a softer spectrum and lower $T_B$. Thus, the $T_B$--$\delta$ trend {\gf might be viewed} as an observational projection of this acceleration--transport balance.

The SHS time profiles then show individual bursts evolving along this same trend. This provides a crucial constraint on the acceleration and transport models, which must explain not only the temporal SHS evolution in single events, but also the reproducible $T_B$--$\delta$ locus across events. Although the present data do not uniquely identify the acceleration mechanism, the common locus suggests a broadly shared dependence of the electron spectral slope on the effective acceleration/transport balance in the \mw-emitting source. Events or phases that depart from this clean SHS behavior, especially those with hard spectra persisting into the decay phase, may indicate stronger contributions from trapping, energy-dependent escape, or delayed transport effects \citep[e.g.,][]{1998ARA&A..36..131B,2000ApJ...531.1109L,2008ApJ...683.1180G,2023ApJ...953..174F}.

\begin{figure*}
\centering
\includegraphics[width=0.9\linewidth]{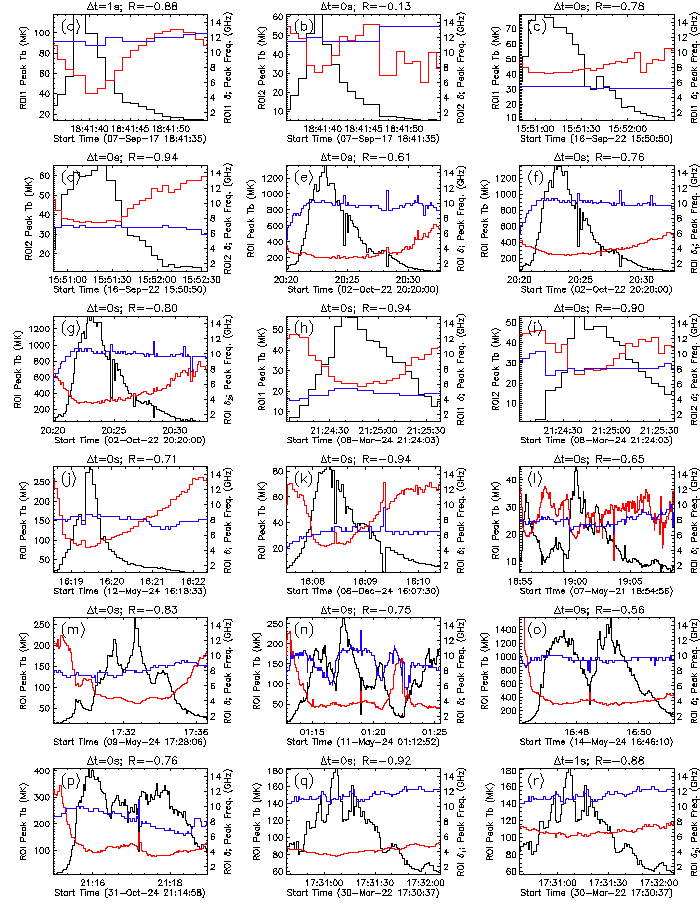}
    \caption{Lag-correlation analysis between the median values of the spectral slope (red) and peak brightness temperature $T_B$ (black), computed over each selected ROI. The corresponding median spectral peak frequencies are shown in blue. The highest cross-correlation coefficient and the corresponding time delay are shown in the panel titles.
    \label{Fig:lag_corr}}
\end{figure*}

We also investigated the lag-correlations and potential delay between the spectral index and peak flux or peak brightness temperature. Like in the case of Fig.\,\ref{Fig:HID}, we found that $T_B$ is typically better correlated with $\delta$ than $S_{\rm peak}$, which to a certain extent may be due to the time variation of the spectral peak frequency during the flare. Figure\,\ref{Fig:lag_corr}, complementary to the set of figures \ref{Fig:delta_2017-Sep-07}--\ref{Fig:delta_2022-03-30}, displays the time evolution of the median brightness temperature $T_B$, $\delta$, and $f_{\rm peak}$, along with the lag between $T_B$ and $\delta$ and the largest cross-correlation coefficient between these values.  In almost all cases, the maximum cross-correlation occurs at zero lag. Only two cases, panels (a) and (r), show a one-bin lag, and both correspond to events analyzed with a 1~s integration time. Therefore, within the cadences ($t_{\rm int}$) used for the presented analysis, see Table\,\ref{table_combined_corr}, the evolution of $T_B$ and $\delta$ do not reveal any noticeable time delay. The cross-correlation coefficients are highly significant; their absolute values range between $\sim0.6$ and $\sim0.95$. 
This confirms close association between the brightness temperature and the spectral slope of the nonthermal electrons.

The discovered prominent spectral evolution of the \mw-producing, mildly relativistic electrons is in a huge contrast with that deduced from the HXR data. Indeed, although HXR bursts often display a SHS evolution, the range of the corresponding spectral index variation is typically rather modest, e.g., 6--4--6 or so. The SHS evolution of the mildly relativistic electrons reported here is much more prominent, 15--(3-5)--15. It is possible that deca-keV electrons also originate with a comparably steep spectrum at the very early stage of their acceleration in the flare, but they cannot yet be detected at that stage because of insufficient sensitivity of the X-ray instruments. At a ``developed'' rise phase, however, the evolutionary patterns of the deca-keV and mildly relativistic electrons are notably different \citep{2021ApJ...908L..55C}: the lower-energy, HXR-producing, electrons show almost no spectral evolution, while the mildly relativistic, \mw-producing, electrons display a highly prominent spectral evolution. This implies that the particle acceleration process ``propagates'' in time from lower to higher energies, which typically takes dozens of seconds in agreement with models of stochastic acceleration \citep[e.g., ][]{Byk_Fl_2009}.

\begin{acknowledgments}
EOVSA was designed, built, and is now operated by the New Jersey Institute of Technology (NJIT) as a community facility. EOVSA operations are supported by NSF grant AGS-2436999 to NJIT.
This work was supported in part by NSF grant 
AGS-2425102  
and NASA grants
80NSSC23K0090 
and 80NSSC21K0520 
to NJIT and CSTR core funding. TK was supported by DFG grant eBer-24-58553.
.
\end{acknowledgments}

\end{document}